\documentclass[apj]{emulateapj} 

\usepackage{bm}
\usepackage{graphicx}
\usepackage{epsf}
\usepackage{graphics}

\def\vec{\mathbf}

\def\mpc{\,h^{-1}{\rm Mpc}}
\def\kpc{\,h^{-1}{\rm kpc}}

\def\msun{\,h^{-1}{\rm M}_\odot}
\newcommand{\rmag}{\>^{0.1}{\rm M}_r-5\log h}
\newcommand{\kms}{\>{\rm km}\,{\rm s}^{-1}}

\makeatletter

\newcommand{\Rmnum}[1]{\expandafter\@slowromancap\romannumeral #1@}
\makeatother

\shorttitle{Genus statistics of cosmic large-scale structure}

\shortauthors{Zhang, Springel \& Yang}

\begin{document}


\title{Genus statistics using the Delaunay tessellation field estimation
  method:\\ (I) tests with the Millennium Simulation and the SDSS DR7}

\author{Youcai Zhang\altaffilmark{1,3},  Volker Springel\altaffilmark{2,4}, Xiaohu
  Yang\altaffilmark{1} }

\altaffiltext{1}{Key Laboratory for Research in Galaxies and Cosmology,
  Shanghai Astronomical Observatory; the Partner Group of MPA; Nandan Road 80,
  Shanghai 200030, China; E-mail: yczhang@shao.ac.cn}

\altaffiltext{2}{Max-Planck-Institut f\"ur Astrophysik,
  Karl-Schwarzschild-Strasse 1, 85748 Garching, Germany}

\altaffiltext{3}{Graduate School of the Chinese Academy of Sciences, 19A,
  Yuquan Road, Beijing, China}

\altaffiltext{4}{Heidelberg Institute for Theoretical Studies,
  Schloss-Wolfsbrunnenweg 35, 69118 Heidelberg, Germany}

\begin{abstract} We  study the topology  of cosmic large-scale structure  
through the genus statistics, using galaxy catalogues generated from the
Millennium Simulation and observational data from the latest Sloan Digital Sky
Survey Data Release (SDSS DR7).  We introduce a new method for constructing
galaxy density fields and for measuring the genus statistics of its isodensity
surfaces. It is based on a Delaunay tessellation field estimation (DTFE)
technique that allows the definition of a piece-wise continuous density field
and the exact computation of the topology of its polygonal isodensity
contours, without introducing any free numerical parameter.  Besides this new
approach, we also employ the traditional approaches of smoothing the galaxy
distribution with a Gaussian of fixed width, or by adaptively smoothing with a
kernel that encloses a constant number of neighboring galaxies. Our results
show that the Delaunay-based method extracts the largest amount of topological
information. Unlike the traditional approach for genus statistics, it is able
to discriminate between the different theoretical galaxy catalogues analyzed
here, both in real space and in redshift space, even though they are based on
the same underlying simulation model. In particular, the DTFE approach detects
with high confidence a discrepancy of one of the semi-analytic models studied
here compared with the SDSS data, while the other models are found to be
consistent.
\end{abstract}
\section{INTRODUCTION}\label{sec_intro}

According to the standard cosmological model, the large-scale structure in the
present Universe has formed as a result of gravitational amplification of
primordial density fluctuations which were presumably seeded by quantum
fluctuations during the early phases of cosmic inflation \citep{Guth1981,
  Guth1982}. The statistical properties of the primordial fluctuations are
well constrained by cosmic microwave background (CMB) observations, as for
example measured by the {\it{Wilkinson Microwave Anisotropy Probe (WMAP)}}
\citep{Dunkley2009}. If the Universe is indeed dominated by cold dark matter,
the statistics of the initial density field are thought to be well preserved
in the large-scale distribution of galaxies, as mapped, e.g., by the Sloan
Digital Sky Survey (SDSS) or the 2dF Galaxy Redshift Survey \citep[2dFGRS,] []
{Colless2001}.  However, it remains an important task to test whether the
observed galaxy distribution is indeed consistent with the expectations for
the prevailing $\Lambda$CDM cosmology. The latter can be accurately obtained
by evolving the initial conditions constrained by the CMB forward in time with
the ordinary laws of physics \citep[e.g.][]{Springel2006}.

Most studies of the large-scale distribution of matter and galaxies are based
on low-order two-point statistics, i.e.~the two-point correlation function and
its Fourier transform, the power spectrum \citep{Peebles1980}. While these
provide the most basic characterization of the statistics of a set of discrete
points, they only fully characterize Gaussian random fields.  In order to
detect deviations from Gaussianity and the higher-order correlations that
develop as a result of non-linear growth of structure even from Gaussian
initial conditions, different statistical measures are required. For example,
this can be done in terms of the one-point density distribution
\citep{Kofman1994}, or through the three-point correlation function and the
bispectrum.  An interesting alternative are more direct measures of the
geometry and topology of cosmic large-scale structure, such as shape
statistics \citep{Dave1997}, Minkowski functionals \citep{Mecke1994}, or the
genus statistics \citep{Gott1986}.  These morphological measures of
large-scale structure are sensitive to higher-order correlations and provide
important descriptive statistics of the cosmic web.

In this paper, we focus on the genus statistic first proposed by
\citet{Gott1986}.  It has been widely applied during the past two decades
\citep{Gott1989, Park1992, Moore1992, Rhoads1994, Vogeley1994, Canavezes1998,
  Springel1998, Hikage2002, Hikage2003, Gott2008, Gott2009, James2009}, in
particular due to its sensitivity to non-Gaussianity.  The genus measures the
topology of isodensity surfaces of a smoothed mass density field.  Therefore,
it is sensitive to global aspects of the density maps. It can test directly
whether the geometry of the observed cosmic web is consistent with theoretical
expectations for the $\Lambda$CDM cosmology.

However,   the  genus  statistics   is  strongly   affected  by   the  density
reconstruction  techniques  that are  required  to  allow  a consideration  of
isodensity contours in  the first place. The most  commonly employed technique
for this purpose uses simple Gaussian  smoothing of the point set with a fixed
kernel size. However, the need  to avoid severe under-sampling in void regions
usually forces one to significantly over-smooth the strongly clustered regions
of the cosmic  web, like the nodes and filaments.  A spatially adaptive kernel
in  real space  may therefore  be a  better choice,  but just  like  the fixed
smoothing,  it  also  introduces  an  unwanted numerical  parameter  into  the
analysis,  namely the  number  of  neighbors used  for  defining the  adaptive
kernel.

In this  work, we therefore propose a  novel method that provides  for a genus
analysis  without  any free  parameter  and  with  minimal smoothing,  thereby
allowing an extraction  of the maximum amount of  topological information from
the point set.  The new approach is based on a three-dimensional density field
reconstruction based  on the  Delaunay Tessellation Field  Estimator technique
\citep{Schaap2000,  Pelupessy2003}.    This  method  allows   the  unambiguous
definition of a  continuous, piece-wise linear density field  on a tetrahedral
mesh, constructed directly  from the coordinates of the  point set. This field
is   consistent,  i.e.~its   volume  integral   reproduces  the   total  mass.
Furthermore, isodensity contours can be  specified exactly in this field; they
become polygonal surfaces for which the genus can also be calculated exactly.

We apply both our new method as well as the traditional approaches of fixed
and adaptive smoothing to a number of different galaxy formation models
constructed from the Millennium Simulation. We also compare genus results
obtained for volume-limited galaxy samples constructed from the SDSS data
release 7 to matching mock surveys we created for the theoretical galaxy
catalogues. This yields an important test of the consistency between the
topology of the observed galaxy distribution and the theoretical models, as
well as direct information about the improvement in discriminative power made
possible by our new method for measuring the genus.

This paper is organized as follows. In \S\ref{sec_method} we briefly describe
the genus statistics, and introduce the different density reconstruction
methods we use in this study.  In \S\ref{sec_data}, we describe the data that
are used to perform our genus analysis: the galaxy catalogues constructed from
the Millennium Simulation, the SDSS observations, and the mock galaxy redshift
surveys we construct for comparison.  \S\ref{sec_results} presents our results
for the genus curves of the observational and simulated data.  Finally, we
discuss our findings in \S\ref{sec_conclusions}.

\section{GENUS STATISTICS}\label{sec_method}

The  genus   is  a   measure  of  the   topology  of  a   surface.   Following
\citet{Gott1986}, we define the genus as
\begin{eqnarray}
G&=&{\rm number \ of \ holes} \nonumber \\ &-&{\rm number
  \ of \ isolated \ regions},
\end{eqnarray} 
which differs from the usual mathematical definition of the genus, $g_{\rm
  math} = G + 1$, by a constant offset of 1.  In our definition, an isolated
sphere has a genus of $-1$, and a torus has a genus of $0$.  In the analysis
of cosmic large-scale structure, it is customary to analyze the genus of
isodensity surfaces as a function of density threshold for a density field
that is suitably created from a tracer distribution of discrete points
(e.g. galaxies).  The actual measurement of the genus can be carried out by
means of the Gauss-Bonnet theorem,
\begin{equation}
G= - \frac{1}{4\pi} \int \kappa\, {\rm d}{\cal A} ,
\end{equation} 
which relates the genus to the integral of the Gaussian curvature,
$\kappa = 1/(r_1 r_2)$, over the surface. Here $r_1$ and $r_2$ are the
principal radii of curvature of the surface at the integration point.

Interestingly, for a Gaussian random density field, the genus per unit volume,
$g(\nu)=G(\nu)/V$, is given by an analytic formula \citep{Hamilton1986},
\begin{equation}\label{GaussianRandom}
g(\nu) = A(1-{\nu}^2)\exp\left( - \frac {{\nu}^2}{2} \right),
\end{equation} 
where the amplitude 
\begin{equation}
A = \frac {1}{2{\pi}^2} {\left( \frac {\langle k^2 \rangle} {3} \right)}^{3/2}
\end{equation}
depends only on the second moment 
\begin{equation}
\langle k^2 \rangle = \frac {\int k^2 P(k) W(k) {\rm d}^3 k } {\int P(k) W(k)
{\rm d}^3 k}
\end{equation} 
of the shape of the (smoothed)  power spectrum $P(k)$. Here $W(k)$ denotes the
Fourier transform of the smoothing kernel.  The dimensionless parameter $\nu =
\delta_t/\sigma$ encodes  the density  threshold $\delta_t$ of  the isodensity
surface that is considered, expressed  in units of the rms dispersion $\sigma$
of the field.  The density  field itself is expressed as dimensionless density
fluctuation field, $\delta(\vec{x}) = \rho(\vec{x})/\left<\rho\right> - 1$.

For general density fields, it is more  useful to define $\nu$ in terms of the
fraction  $f$ of  the   volume  above  the density  threshold, through  the
equation
\begin{equation}\label{vfraction}
f(\nu)= \frac {1}{2} \rm {erfc} \left(\frac {\nu}{\sqrt {2}}\right), 
\end{equation} 
where ${\rm erfc}(x) \equiv 2{\pi^{-1/2}} \int_{x}^{\infty}{e^{-t^2}{\rm d}t}$
is the complementary error function. The  $\nu = 0$ contour corresponds to the
median volume fraction contour ($f=50\%$).  We will employ this definition for
labeling our  density contours, but  note that for  a Gaussian field  one then
still has $\nu = \delta_t/\sigma$.

Deviations of the shape $g(\nu)$ of a measured genus curve from the form
of Eqn.~(\ref{GaussianRandom}) can be interpreted as a measure of
non-Gaussianity, or equivalently, of the cumulative impact of
higher-order correlations.  Our procedure for quantifying such
deviations is to first find the best-fitting genus curve $g_{\rm
  fit}(\nu)$ expected for Gaussian random phases through a least-squares
fit of Eqn.~(\ref{GaussianRandom}) to the measured curve. This yields,
in particular, the {\em amplitude} of the measured genus curve.

Next, one can define different  meta-statistics to quantify the differences of
the measured  genus curve  relative to the  Gaussian shape.  For  example, the
shift parameter $\Delta\nu$ has been defined by \citet{Park1992} as
\begin{equation}\label{delta_nu}
\Delta\nu = \frac {\int_{-1}^1 {g(\nu)\,\nu \,{\rm d}\nu}}
{\int_{-1}^1 {g_{\rm fit}\, {\rm d}\nu}},
\end{equation} 
which measures the horizontal shift of the central portion of the genus
curve.  The Gaussian random phase curve (Eqn.~\ref{GaussianRandom}) has
$\Delta\nu = 0$.  A negative value of $\Delta\nu$ is sometimes called a
``meatball shift'' because it signifies a greater prominence of isolated
over-densities in the density field, while a void dominated field leads
to $\Delta\nu > 0$, called a ``bubble shift''.

Since the two negative extrema of  the genus curve represent a measure for the
frequency of isolated over- and  under-densities, the abundance of clusters and
voids relative to that expected from  the best-fit Gaussian genus curve can be
quantified by two parameters $A_C$ and $A_V$ \citep{Park2005a}, defined as
\begin{equation}\label{A_C}
A_C = \frac {\int_{1.2}^{2.2} {g(\nu)\, {\rm d}\nu}}
{\int_{1.2}^{2.2} {g_{\rm fit}(\nu)\,{\rm d}\nu}},
\end{equation}
and
\begin{equation}\label{A_V}
A_V = \frac {\int_{-2.2}^{-1.2} {g(\nu)\, {\rm d}\nu}}
{\int_{-2.2}^{-1.2} {g_{\rm fit}(\nu) \,{\rm d}\nu}},
\end{equation} 
where the integration interval of $A_C$ ($A_V$) is roughly centered at
$\nu = \sqrt{3}$ and $\nu = - \sqrt {3}$, respectively, which are the
locations of the minima in the best-fit Gaussian genus curve. At these
points the sensitivity to the number of clusters and voids is
greatest. $A_{C,V} < 1$ ($A_{C,V} > 1$) means that fewer (more) isolated
clusters or voids are observed than those expected from the best-fit
Gaussian curve.

In  principle, additional genus  meta-statistics are  conceivable, and  one may
even  use  principal  components  analysis  to determine  the  most  sensitive
measures for shape distortions  of the genus curve \citep{Springel1998}. Also,
the amplitude  of the genus curve  relative to the  genus of a field  with the
same  power  spectrum \citep{Canavezes1998}  can  be  used  as a  quantitative
measure for  the cumulative effect  of higher order correlations.  However, in
this  study,  we  will  restrict  ourselves to  the  simple  genus  statistics
described above.

\subsection{Fixed Smoothing}

As is  clear from  the above,  a given discrete  point set  first needs  to be
transformed to a continuous density field to allow the genus analysis. The simplest
approach for  this is to smooth  the point set  with a Gaussian kernel  of the
form
\begin{equation}\label{window_function}
W(r) = \frac {1} {(2\pi\lambda^2)^{3/2}} \exp \left( - \frac {r^2}{2\lambda^2}
\right), 
\end{equation} 
where $\lambda$ is  a fixed smoothing length. In  practical terms, a Cartesian
grid with spacing  $d$ can  be used to represent  the density field, where
the grid spacing $d$  needs to be smaller than $\lambda$ by  a factor of a few
to provide adequate  sampling. Sampling constraints also impose  a lower bound
on  reasonable values  for $\lambda$,  which should  be  chosen (considerably)
larger than the mean spacing of  points in the sample, otherwise the estimated
density field will be dominated by noise. In practice, we choose $\lambda$ to
be within $5-10\mpc$.

The actual measurement of the genus can then be carried out on the Cartesian
grid in the following way.  For a given prescribed density threshold, one
considers the surface of all cell faces that lie between pairs of cells on
opposite sides of the density threshold. This surface is then an approximation
to the true isodensity contour of the underlying density field. In fact, the
topology of the discretized surface, which is composed of squares, will be the
same as that of the smooth true isodensity contour (modulo very small changes
if the grid spacing is not fine enough), except that all the curvature of the
real surface is now compressed into the corners of the discretized
contour. All the curvature is then represented by a set of Dirac delta
functions at these corners, each giving an integrated curvature equal to the
angle deficit relative to 360 degrees for the sum of the angles between the
incident edges. An application of the Gauss-Bonnet theorem to measure the
genus of the surface then effectively reduces to summing up the angle deficits
at all vertices of the contour.  This approach is exploited by the {\small
  CONTOUR} algorithm \citep{Weinberg1988}, which we have reimplemented in a C
code for the analysis of the smoothed density fields in this paper.

\subsection{Adaptive Smoothing}

A serious disadvantage of the fixed smoothing technique is that a single fixed
smoothing  length   is  not  well   adjusted  to  highly   clustered  particle
distributions, like the ones we expect  for galaxies of the cosmic web. If the
smoothing is  chosen sufficiently large  to avoid under-sampling of  the voids,
the filaments and clusters will typically be over-smoothed, limiting the amount
of information that can be extracted.

A better match to the spatially varying sampling density can be obtained if an
adaptive smoothing kernel is used, where  the size of the kernel is determined
by the  local sampling density.  In  our study, we employ  the adaptive kernel
estimation  techniques  that  is  used  in the  well-known  smoothed  particle
hydrodynamics (SPH)  approach. For the  smoothing kernel, we adopt  the common
choice in SPH, a spherically symmetric spline of the form \citep{Monaghan1985}
\begin{equation}\label{SPH_kernel}
W(r;h) = \frac {8} {\pi h^3}
         \left \{
         \begin{array}{ll}
           1-6{\left( \frac {r}{h} \right)}^2 + 6{\left( \frac {r}{h} \right)}^3 , &
           0 \le \frac {r}{h} \le \frac {1}{2} , \\
           2 {\left( 1 - \frac {r}{h} \right)}^3 , &
           \frac {1}{2} < \frac {r}{h} \le 1, \\
           0 , & \frac {r}{h} > 1 ,
         \end{array} 
         \right. 
\end{equation} 
where the adaptive smoothing length $h({\vec x})$ at each point ${\vec
  x}$ is defined as the distance to the $N_{\rm sph}$-th nearest neighbor such that
the number of particles inside the smoothing radius $h({\vec x})$ is
equal to a constant value $N_{\rm sph}$. We then compute the density
field as
\begin{equation}
\rho(\vec x) = \sum_{i=1}^N m_i W({\vec r_i} - \vec x;h(\vec x)),  
\end{equation} 
where  $m_i$ is  the mass  of the  $i$-th particle.   This corresponds  to the
so-called    {\em   gather}    approach    to   define    a   density    field
\citep{Hernquist1989}.

Again, we shall  use a finely spaced Cartesian grid  to represent the smoothed
field. We  note however that  it is here  considerably more difficult  to make
this mesh fine enough everywhere.   In fact, for clustered distributions, some
unwanted over-smoothing  of the densest regions  due to a too  coarse mesh can
usually not be avoided.

\begin{figure*}
\begin{center}
\resizebox{16cm}{!}{\includegraphics{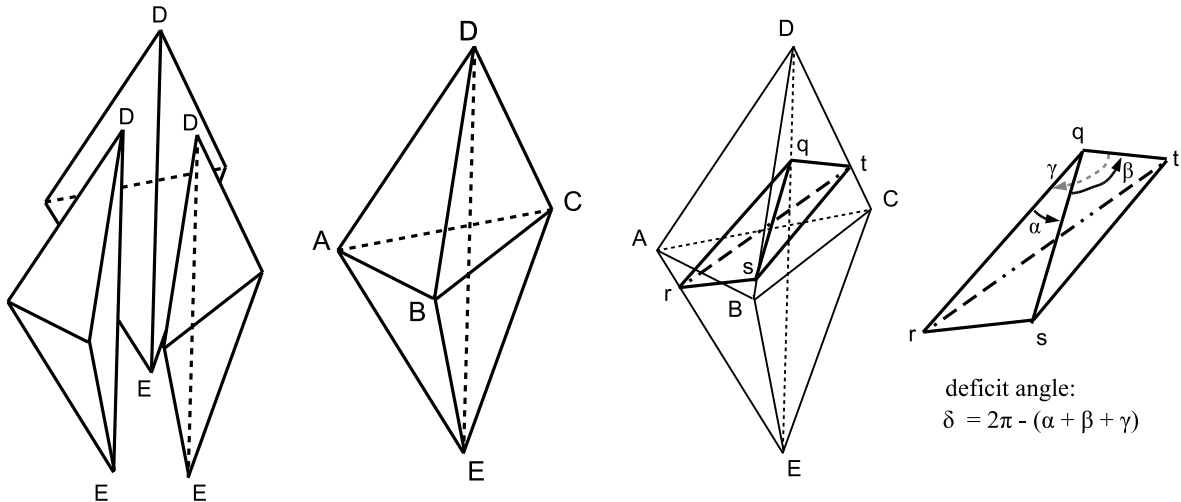}}
\end{center}
\caption{Sketch of the curvature calculation for isodensity surfaces
  constructed with the Delaunay Tessellation Field Estimator (DTFE)
  technique. The drawing on the left illustrates the three
    Delaunay tetrahedra that share a common edge DE in the
    tessellation, where for clarity the tetrahedra have been displaced
    from each other. The second drawing shows the real geometry of the
    involved subgroup of 5 points (galaxies) in the point set.  We
    assume that the density threshold $\rho_t$ of the contour lies
    between the densities $\rho_{\rm D}$ and $\rho_{\rm E}$ estimated
    for the points D and E.  We can then compute the point q in which
    the isodensity contour intersects the edge DE. Similarly, the
    isodensity contour intersects the outer edge DA-AE in point r, the
    lines DB-BE in point s, and DC-CE in point t (see third drawing
    from left). Finally, we can compute the angles $\alpha$, $\beta$
  and $\gamma$ enclosed by the edges of the isodensity contour that
  are incident on point q (rightmost drawing). This yields the angle
  deficit at this corner of the isodensity contour, providing a
  measurement of the integrated curvature.}
\label{fig:sketch}
\end{figure*}

\subsection{Genus calculation through a Delaunay tessellation}

The  above limitations  of the  fixed and  adaptive smoothing  techniques have
motivated us to develop a more  general approach for measuring the topology of
a point set. The  new method we propose below does not  rely on any additional
parameter,  it is  free  of grid-spacing  limitations,  and at  the same  time
extracts the maximum topological information from the point set.

In mathematics and computational geometry, the Delaunay tessellation for a set
of points is the uniquely defined and volume-covering tessellation of mutually
disjoint tetrahedra, in which no  circumsphere of any tetrahedron contains one
of the points in its  interior \citep{Delaunay1934, Okabe2000, Illian2008}. 
Connecting the centers of
the  circumscribed spheres  produces the  Voronoi tessellation,  which  is the
topological dual of the Delaunay tessellation.

There are different possibilities to use the Delaunay or Voronoi tessellations
for density  estimates. For example, one  may use the  Voronoi volumes $V^{\rm
Vor}_i$,  and  simply  estimate  the  density  within  each  Voronoi  cell  as
$\rho_i^{\rm Vor} = \frac {m_i}  {V^{\rm Vor}_i}$. Another possibility lies in
defining the density as
\begin{equation}
\rho_i = \frac {4\,m_i} {W_{i}},
\label{eqnConDel}
\end{equation} 
where now $W_{i}$ is the volume of the contiguous Delaunay region around point
$i$,  i.e.~the  sum $W_i  =\sum_j  V^{\rm Del}_{ij}$  of  the  volumes of  all
Delaunay tetrahedra  that have $i$ as  one of their vertices.   Note that each
Delaunay tetrahedron is contributing to the contiguous Delaunay region of four
points,  hence the  multiplication  by 4  in  Eqn.~(\ref{eqnConDel}).  In  the
Delaunay      Tessellation      Field      Estimator     (DTFE)      technique
\citep{Schaap2000,Pelupessy2003},  one estimates densities  for each  point in
this way,  and then linearly extends  the density estimates at  the corners of
each tetrahedron to  a volume filling density field.  To  this end, one simply
uses tri-linear  interpolation of  the density values  at the corners  of each
tetrahedron,  in  the basis  of  the three  principal  vectors  that span  the
tetrahedron. This  creates a continuous,  piece-wise linear density  field, in
which the  gradient within  each tetrahedron is  constant. Furthermore,  it is
easy to see that  the volume integral of this field reproduces  the sum of the
particle masses exactly.

The  above  DTFE  method  allows  a  unique,  parameter-free  construction  of
polygonal  isodensity  contours  for  a  given  set  of  points,  without  any
restriction   on   dynamic   range,   and  without   requiring   a   smoothing
procedure.  Furthermore,  we  can  readily  measure  the  integrated  Gaussian
curvature (and hence the genus)  of this surface by generalizing the technique
employed for Cartesian grids.  This is because again all the curvature will be
compressed   into   the   corners/vertices   of   the   polygonal   isodensity
surface. These points  all lie on edges of  the Delaunay tessellation, because
only there three or more tetrahedra with different gradients can meet.

An   example   for  the   geometrical   situation   is   sketched  in   Figure
~\ref{fig:sketch}. For  every edge in  the Delaunay tetrahedralization  we can
readily  decide whether or  not it  is intersected  by the  isodensity contour
corresponding to  a prescribed density  threshold $\rho_t$. This  happens when
the densities of the  two endpoints of the edge lie on  different sides of the
density threshold.  Provided  this is the case, we  can straightforwardly find
the intersection point  on the edge which has density  $\rho_t$. Next, we have
to determine  the deficit angle around this  point. To this end,  we visit all
tetrahedra that share the given edge  and find the points on their outer edges
that have density $\rho_t$. Based on simple geometrical operations we can then
determine  the  total  angle  deficit  around the  intersection  point,  which
corresponds  to the  integrated curvature  at  this corner  of the  isodensity
contour.   Summing  these  angles  over   all  Delaunay  edges  that  have  an
intersection point then yields the genus of the surface.

We can also obtain an exact measurement of the enclosed volume fraction within
the isodensity  contour. To this end,  we simply loop over  all tetrahedra and
check whether any of their points lie below or above the density threshold. If
all four  points lie  below $\rho_t$,  the full volume  of the  tetrahedron is
counted in the  enclosed volume fraction.  If one  corner lies below $\rho_t$,
then the intersections on the three incident edges can be calculated, yielding
together with the low corner a  tetrahedral volume that is counted towards the
volume  fraction. Similarly,  if three  points  are below  $\rho_t$, then  the
volume complement for the corner that  lies above the density threshold can be
readily computed.  Only in the case where  two points lie below  and two above
$\rho_t$, a bit of more work  is required. Here the volume below the threshold
can be obtained as a  composite of three suitably defined tetrahedra. Finally,
if all four points lie above the threshold, there is evidently no contribution
to the enclosed volume fraction.
 
In our data analysis below, we will  try out this method for the first time in
the  topological analysis of  cosmic large-scale  structure. To  construct the
Delaunay  mesh, we  employ the  tessellation  engine of  the parallel  {\small
AREPO} code \citep{Springel2009}.

\begin{figure*}
\begin{center}
\resizebox{8cm}{!}{\includegraphics{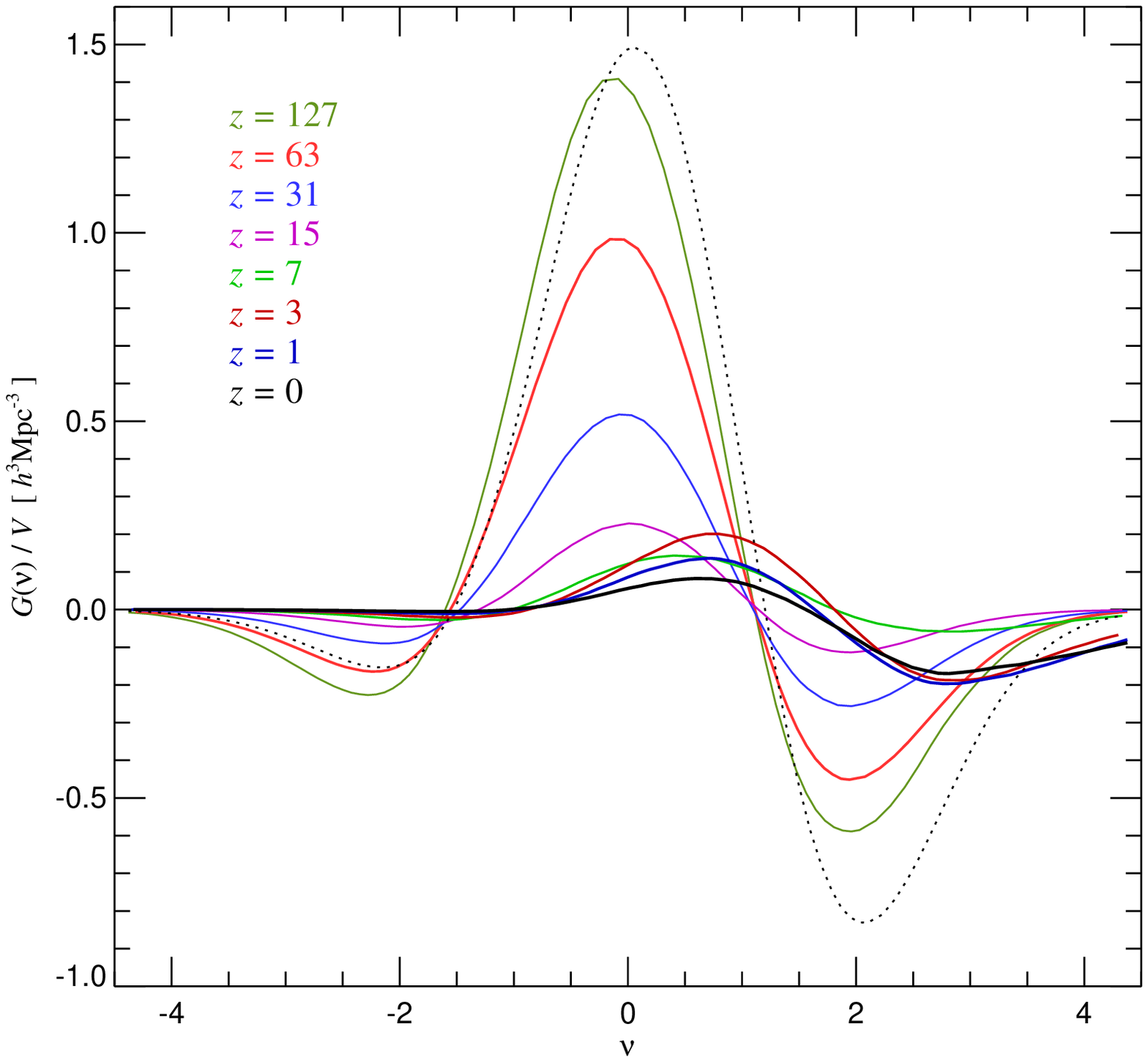}}%
\resizebox{8cm}{!}{\includegraphics{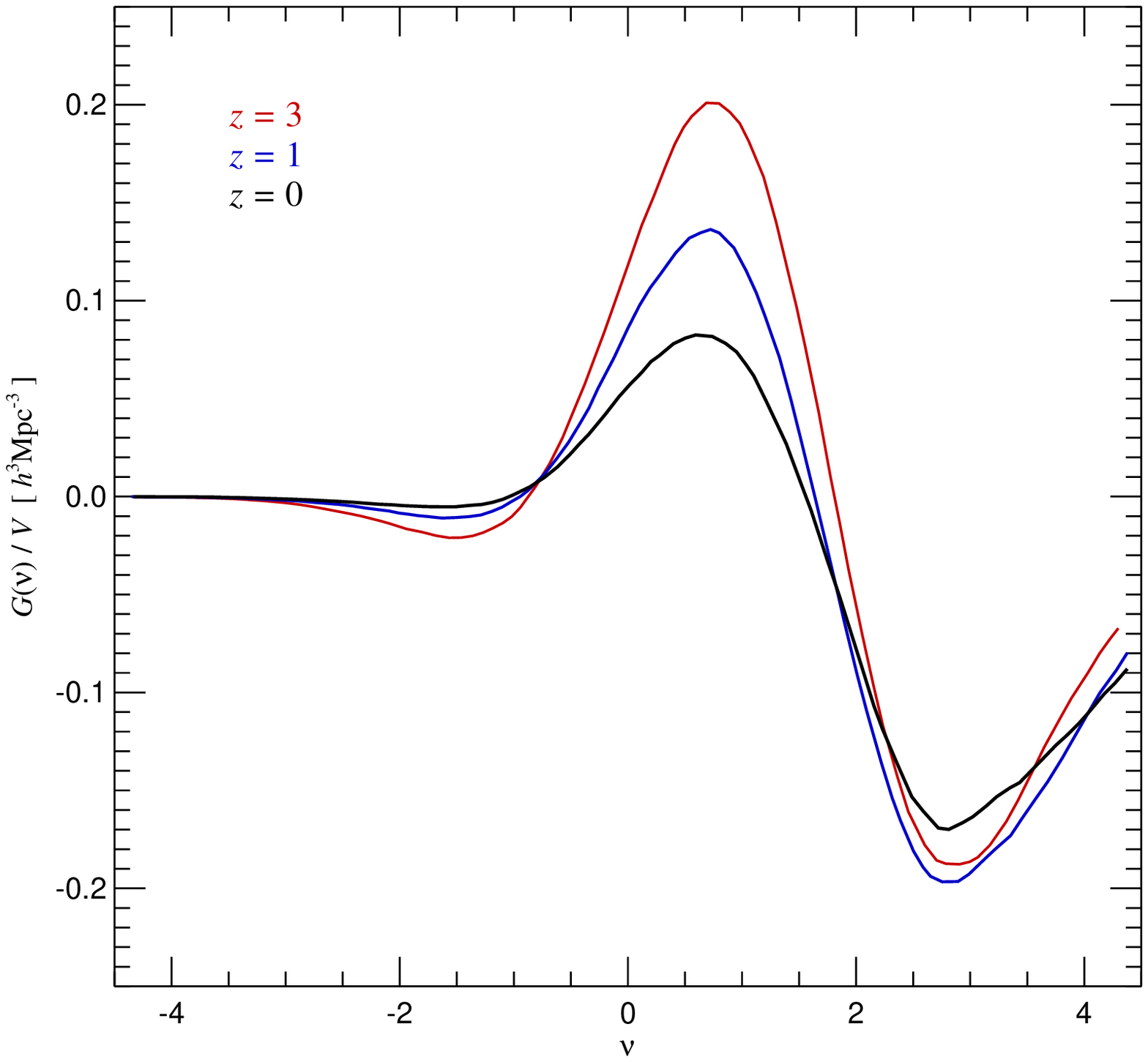}}%
\end{center}
\caption{Genus curves calculated with the DTFE technique at different times
  for a dark matter N-body simulation of the $\Lambda$CDM cosmology.  The
  panel on the left illustrates the substantial modification of the shape of
  the genus curve as a function of time, first through mildly non-linear
  evolution, and finally through the strong clustering in the non-linear
  regime.  The dotted line shows the curve for a random distribution of an
  equal number of points ($128^3$) in the box of size $V = (50\,h^{-1}{\rm
    Mpc})^3$.  The panel on the right repeats the low-redshift measurements,
  for clarity. In this strongly clustered regime, the genus shows a marked
  ``meatball'' shift, and a very strong asymmetry.}
\label{fig:nbody}
\end{figure*}

Before we apply this new approach to galaxy data, it is useful to illustrate
its response to clustering with the help of an N-body simulation. To this end,
we have run a small collisionless N-body simulation with $128^3$ particles in
a periodic box of size $L=50\,h^{-1}{\rm Mpc}$ on a side, using the
cosmological parameters of the Millennium Simulation.  In Figure
\ref{fig:nbody}, we show eight genus curves measured at different times
between the starting redshift $z=127$ and the final epoch of $z=0$. First,
during the mildly non-linear evolution, the genus amplitude declines, due to
the development of phase correlations \citep{Springel1998}.  After $z\sim 7$,
strong non-linear evolution sets in, and the number density of virialized
halos rapidly increases. As a result, the genus curve develops a ``meatball
shift'', and an ever large asymmetry between the minima on the $\nu< 0$ and
$\nu > 0$ sides. Interestingly, the maximum of the genus curve first begins to
increase in this non-linear phase, but then declines again at late times. The
latter is probably related to the slowly declining total number of halos at
late times, as they aggregate into ever larger structures, and the
thinning-out of the cosmic web.  For comparison, we also include in
Figure~\ref{fig:nbody} the genus curve for a random distribution of an equal
number of points, shown as a dashed line, which yields a very different shape
compared to the N-body simulation at all times. Overall, it is therefore clear
that the shape of the genus curve measured with the DTFE method encodes a
wealth of interesting information about the clustering pattern which is not
readily accessible by other statistical measures.

\section{DATA}\label{sec_data}

\begin{figure}
\begin{center}
\resizebox{8cm}{!}{\includegraphics{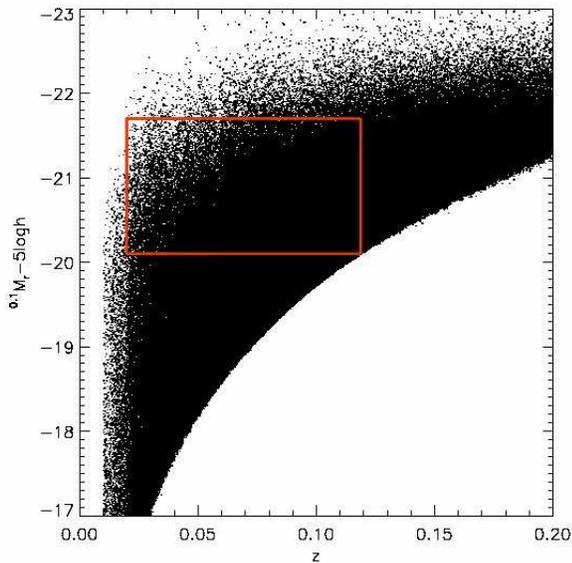}}
\end{center}
\caption{SDSS  galaxies in redshift-absolute  magnitude space.   The rectangle
  indicates  the  boundary definition  we  adopted  to  define volume  limited
  subsamples.}
\label{fig:region}
\end{figure}

\subsection{Galaxy catalogues based on the Millennium Simulation}
\label{sec:catalogues}

The simulation analyzed in this study is the so called ``Millennium
Simulation'' (MS) which evolved $N = 2160^3 \simeq 1.0078 \times 10^{10}$
particles from redshift $z=127$ to the present in a cubic box of length
$L_{\rm box} = 500 \mpc$ on each side \citep{Springel2005b}.  The cosmological
parameters in the Millennium Simulation are $\Omega_{\rm m} = \Omega_{\rm
  dm}+\Omega_{\rm b}=0.25$, $\Omega_{\rm b}=0.045$, $h=0.73$,
$\Omega_\Lambda=0.75$, $n=1$, and $\sigma_8=0.9$, where the Hubble constant is
given as $H_0 = 100 \, h {\rm \, km \, s^{-1} \, Mpc^{-1}}$, $\sigma_8$ is the
$rms$ linear mass fluctuation in a sphere of radius $8 \mpc$ at $z=0$ and $n$
is the spectral index of the primordial power spectrum.  The simulation was
carried out with the massively parallel {\small GADGET-2} code
\citep{Springel2005a}.  Gravitational forces were computed with the TreePM
method, where long-range forces are calculated with a classical particle-mesh
method while short-range forces are determined with a hierarchical tree
approach \citep{Barnes1986}.  The Plummer equivalent gravitational softening
length was $5\kpc$, which can be taken as the spatial resolution of the
simulation.

Dark matter halos and subhalos were identified with the FOF \citep{Davis1985}
and {\small SUBFIND} \citep{Springel2001a} algorithms, respectively. Based on
the halos and subhalos at all output times of the simulation, detailed merging
history trees were constructed, which form the basic input required by
subsequently applied semi-analytic models of galaxy formation.

The first galaxy catalogue we consider is constructed based on the conditional
luminosity function (CLF) approach developed by \citet{Yang2003}. Halos of
different masses are populated with galaxies according to the best fit CLF
parameters listed in \citet{Cacciato2009}, properly converted to the MS
cosmology.  We assign each galaxy an $r$-band absolute magnitude $\rmag$,
which is $K+E$ corrected to redshift $z=0.1$ \citep{Blanton2003}.  Note that
by construction, the galaxy catalogue has a luminosity function that is in
very good agreement with the SDSS observations.  From this catalogue, we
select a sample of galaxies in the absolute magnitude range $-21.7< \rmag
<-20.1$ for further analysis. While the choice of this absolute magnitude
range is somewhat arbitrary, it yields a good compromise of a fairly large
volume and a high number density of galaxies as covered by the SDSS
observations (see Fig.~\ref{fig:region}).  The total number of galaxies and
mean separation in this sample are $665,914$ and $5.73\mpc$, respectively. In
order to assess statistical uncertainties due to cosmic variance, we divide
the galaxy catalogue corresponding to the full simulation box into $8$
subsamples with $L_{\rm box}=250\mpc$ each, and consider the scatter among the
results for the individual sub-samples.  To investigate the dependence of the
genus statistics on galaxy absolute magnitudes, we also construct two
different samples with the absolute magnitude ranges $-19.8< \rmag <-19.2$ and
$-18.8< \rmag <-18.3$, respectively, which have a similar number of galaxies
as in the case of $-21.7< \rmag <-20.1$.

The second galaxy catalogue we study is taken from \citet{Croton2006}
\footnote{The  semi-analytic   galaxy  catalogue  is   publicly  available  at
  http://www.mpa-garching.mpg.de/galform/agnpaper.}.     This    semi-analytic
galaxy  formation model  includes a  total of  about $9$  million  galaxies at
$z=0$.   After converting  the  absolute  magnitudes of  galaxies  by a  $K+E$
correction to  redshift $z=0.1$, we  also construct from this  catalogue three
different  samples in  the same  absolute magnitude  ranges that  we described
above.

Finally, the third theoretical galaxy catalogue used in this paper is that of
\citet{Bower2006}.  The catalogue consists of a total of about $24$ million
galaxies.  Again, after $K+E$ correcting the absolute magnitudes and
converting them to redshift $z=0.1$, we construct three galaxy samples in the
absolute magnitude ranges $-21.7< \rmag <-20.1$, $-19.8< \rmag <-19.2$ and
$-18.8< \rmag <-18.3$, respectively, as before. Note that as discussed in
\citet{Liu2010}, galaxies with given luminosity/stellar masses have quite
different distributions in halos of different masses when the
\citet{Bower2006} and \citet{Delucia2007} models are compared, whereas the
latter is very similar to \citet{Croton2006}.

Note that in each of the absolute magnitude ranges, there are slight
differences in the numbers of galaxies among the three catalogues.  Since the
genus curve can quite strongly depend on the total number of galaxy in
consideration \citep{Protogeros1997}, we randomly down-sample the galaxies
where needed according to the \citet{Bower2006} catalogue within the magnitude
range $-21.7< \rmag <-20.1$ (a few percent less than others), so that all our
samples have exactly the same number of galaxies.

\subsection{Sloan Digital Sky Survey}

The Sloan Digital Sky Survey \citep[SDSS][]{York2000}, one of the most
influential galaxy redshift surveys to date, is a multi-filter imaging and
spectroscopic survey to explore the large-scale distribution of galaxies and
quasars.  Here we make use of the New York University Value-Added Galaxy
Catalogue \citep[NYU-VAGC;][]{Blanton2005}, which is based on the SDSS Data
Release 7 \citep{Abazajian2009}.  DR7 marks the completion of the survey phase
known as SDSS-II. It features a spectroscopy that is now complete over a large
contiguous area of the Northern Galactic cap, closing the gap which was
present in previous data releases.  The continuity over this large area is a
great advancement and critical to the statistics of large-scale structure.
From the NYU-VAGC, we select all galaxies in the Main Galaxy Sample with an
extinction corrected apparent magnitude brighter than $r=17.72$, with
redshifts in the range $0.01 \leq z \leq 0.20$ and with a redshift
completeness ${\cal C}_z > 0.7$. The extracted SDSS galaxy catalogue contains
a total number of $639,555$ galaxies. Note that in this catalogue, a very
small fraction of galaxies have redshifts that are borrowed from the Korea
Institute for Advanced Study (KIAS) Value-Added Galaxy Catalog (VAGC)
\citep{Park2005c, Choi2007, Choi2010}.

  Since the complicated survey geometry may potentially impact the
  genus measurements significantly, we select only galaxies with right
  ascension $\alpha$ and declination $\delta$ in the ranges
  $120^{\circ}< \alpha <240^{\circ}$ and $10^{\circ}< \delta
  <55^{\circ}$, which results in a loss of $\sim44$\% of the galaxies,
  of which however nearly half are not in the coherent region. With
  this selection, the remaining SDSS survey volume will suffer less
  from edge effects, and we have a precise understanding about where
  the edges lie in our genus calculations. We note that for our new
  DTFE method we always need to put the data points into a large
  enclosing box for which formally periodic boundaries are imposed in
  order to facilitate the construction of the Delaunay
  tessellation. We have checked that filling the residual volume of
  this box with a sparse grid of background points or leaving it empty
  has no significant influence on our results, confirming that the
  edge effects are sufficiently small (see also the tests in
  Fig.~\ref{fig:geometry} below).

Furthermore, we construct a volume-limited galaxy sample with absolute
magnitudes $-21.7< \rmag <-20.1$.  As shown in the redshift-absolute magnitude
diagram (Fig.~\ref{fig:region}), this corresponds to the redshift range
$0.0197<z<0.1187$ when the apparent magnitude cut ($10.2<m_r<17.7$) is
applied.  The final volume-limited sample we use based on the SDSS
observations contains $92,662$ galaxies.

\subsection{Mock galaxy redshift surveys}
\label{sec:MGRS}

The end product of the CLF approach and the SAMs considered here is a
large sample of galaxies distributed over the dark matter halos in the
cubic simulation box of the Millennium Simulation.  One approach would
be to compare these galaxy samples {\it directly} with the SDSS data.
However, this ignores the fact that the latter is affected by
observational selection effects, especially the survey geometry and
redshift distortion effects.  To make an ``apples-to-apples'' comparison
with the SDSS data, which is essential especially for the genus
statistics, we construct mock galaxy redshift surveys (MGRSs) for the
three galaxy catalogues we discussed in Section \ref{sec:catalogues}.

Our construction of the MGRS here is similar to that described in \citet
{Yang2004} \citep[see also][]{Li2007}.  First, we stack $3\times 3\times 3$
replicas of the simulation box and place a virtual observer at the center of
the stacked boxes.  Next, we assign each galaxy ($\alpha$, $\delta$)
coordinates and remove the ones that are outside the mocked SDSS survey
region.  For each model galaxy in the survey region, we compute its redshift
(which includes the cosmological redshift due to the universal expansion, the
peculiar velocity, and a $35\kms$ Gaussian line-of-sight velocity dispersion
to mimic the redshift errors in the data), its $r$-band apparent magnitude
(based on the $r$-band luminosity of the galaxy).  We eliminate galaxies that
are fainter than the SDSS apparent magnitude limit, and incorporate the
position dependent incompleteness by randomly eliminating galaxies according
to the completeness factors obtained from the survey masks provided by the
NYU-VAGC. To have a measure of the error on the genus statistics, we construct
6 MGRSs for the CLF galaxy catalogue by rotating the simulation boxes.

Finally, similar to the SDSS observational data, we construct volume-limited
galaxy samples with absolute magnitudes $-21.7< \rmag <-20.1$ and a redshift
range of $0.0197<z<0.1187$.  Again, we downsample the resulting galaxy
catalogues according to the one with least number of galaxies if needed so
that they have the same number of galaxies for the genus measurements.

\begin{figure*}
\begin{center}
\resizebox{18.0cm}{!}{\includegraphics{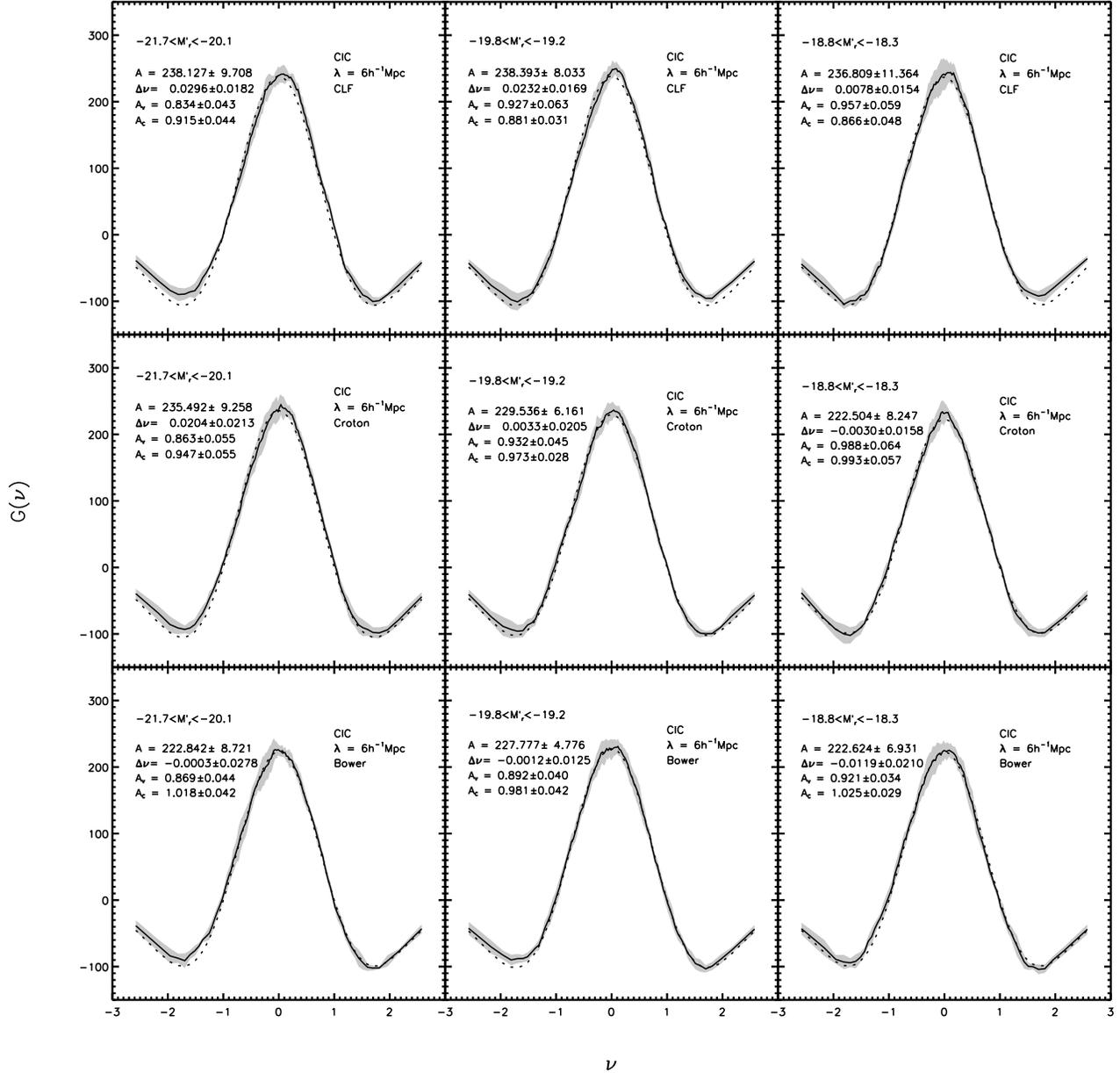}}\vspace*{-0.5cm}
\end{center}
\caption{Genus curves measured for galaxy samples from three different galaxy
  catalogues constructed for the Millennium Simulation, from the upper to
  lower rows as labeled, using the fixed smoothing technique.  The shaded
  region indicates the standard error of the $8$ subsamples in each model,
  while the solid line is their mean. The dotted line is the best-fit genus
  curve for a Gaussian field. The density field has here first been
  constructed on a fine grid using CIC assignment and was then smoothed with a
  Gaussian of smoothing length $\lambda = 6\mpc$.  Results shown in the left,
  middle, and right hand columns are for galaxies in different absolute
  magnitude bins, as labeled, where ${\rm M}'_r=\rmag$ is the $r$-band
  absolute magnitude, $K+E$ corrected to redshift $z=0.1$.  In each panel, we
  include measurements for the amplitude $A$, the horizontal shift
  $\Delta\nu$, and the abundance diagnostics for clusters and void, $A_C$ and
  $A_V$. In each case we cite the mean of our $8$ measurements and their
  standard deviations.  }
\label{fig:cic}
\end{figure*}

\begin{figure*}
\begin{center}
\resizebox{18cm}{!}{\includegraphics{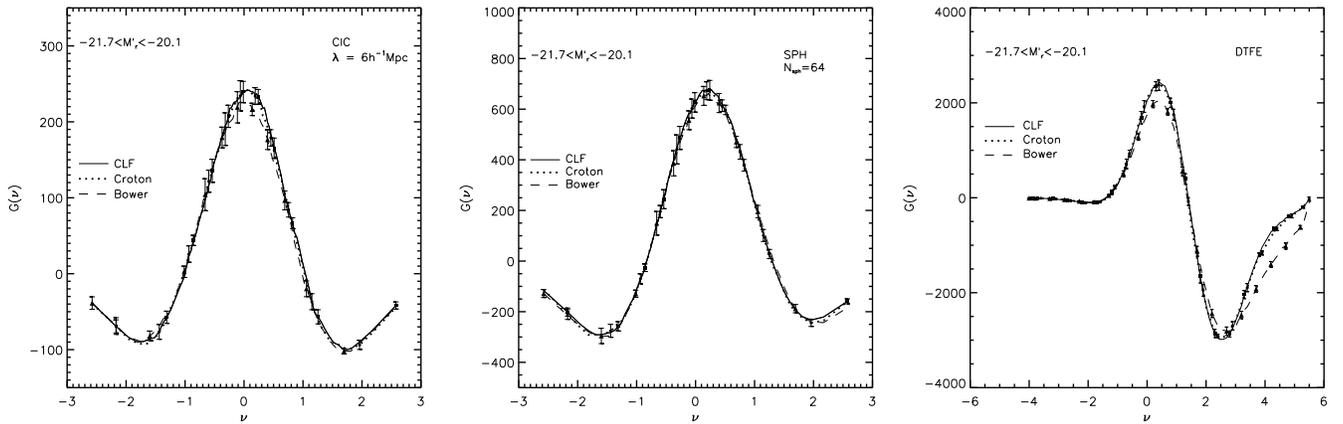}}\vspace*{-0.5cm}
\end{center}
\caption{Similar to Fig.~\ref{fig:cic}, but here we compare results for
  galaxies with $-21.7<\rmag < -20.1$ using three different methods for
  density reconstruction.  In the left panel, the density field was
  constructed using the CIC method with $\lambda=6\mpc$.  In the middle panel,
  the density field was constructed using the SPH adaptive kernel method with
  $N_{\rm sph}=64$ neighbors.  The right panel is based on the DTFE method and
  represents a direct genus measurement of the Delaunay tetrahedralization.
  In each panel, we show results for galaxy subsamples constructed using the
  CLF approach (solid line with squares and error bars), and obtained for the
  semi-analytic models of \citet{Croton2006} (dotted line with circles and
  error bars) and \citet{Bower2006} (dashed line with triangles and error
  bars), respectively.}
\label{fig:method}
\end{figure*}

\begin{figure*}
\begin{center}
\resizebox{18cm}{!}{\includegraphics{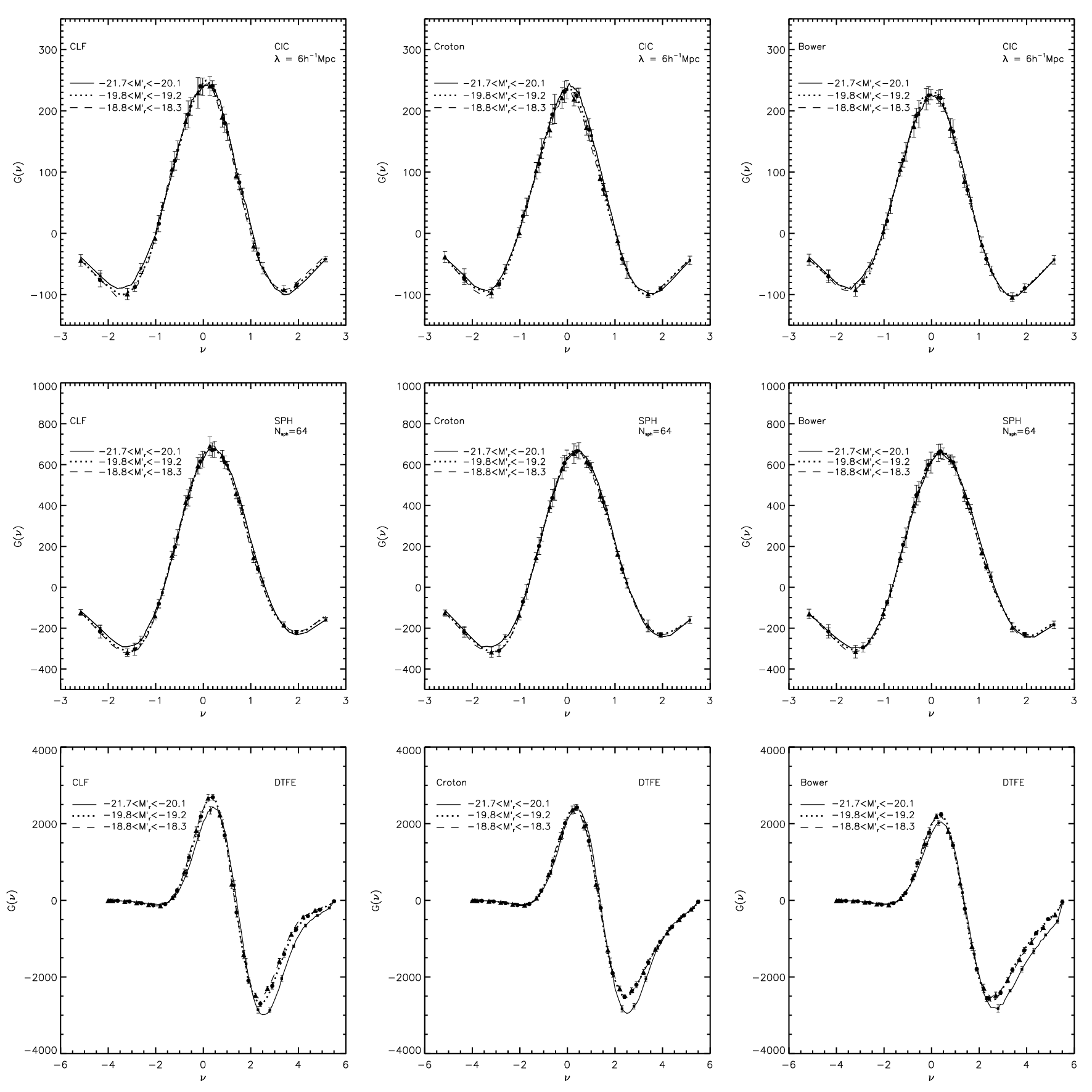}}\vspace*{-0.5cm}
\end{center}
\caption{Comparison of the genus statistics in different absolute magnitude
  bins based on different density reconstruction methods and for different
  galaxy catalogues, as indicated in each panel.}
\label{fig:magnitude}
\end{figure*}

\section{GENUS STATISTICS OF COSMIC LARGE-SCALE STRUCTURE}
\label{sec_results}

Reconstructing a density field from a set of irregularly sampled points is a
key step in measuring the genus curve. In this section, we investigate the
effects of different reconstruction methods for the density field on the genus
and its related statistics. Since we have two types of galaxy catalogues, with
and without SDSS survey selection effects, we present our measurements in two
subsections. 

\subsection{Results for galaxies in real space}

We first carry out genus measurements for the three galaxy catalogues in
the full Millennium Simulation box, where periodical boundary conditions
apply.  Note that we are here measuring the signal in real space, i.e.,
the distribution of galaxies is not affected by the redshift distortion
effects that are present in the SDSS observations.

\subsubsection{Fixed Smoothing Genus Curve}

The mean galaxy separation in all of our galaxy samples based on the
Millennium Simulation is about $5.73\mpc$, therefore we can reasonably safely
apply a Gaussian smoothing length of $6\mpc$.  If the smoothing length is
smaller than $1/\sqrt{2}$ times the mean galaxy separation, the genus curve
tends to show a ``meatball shift'' because the algorithm picks out individual
galaxies as isolated high density regions \citep{Gott1987, Gott1989}.

For each of $8$ subsamples of the three full galaxy catalogues, we first
compute the density field using the cloud-in-cell (CIC) assignment scheme. The
galaxies are assigned onto $64^3$ grids covering the box of $L_{\rm
  box}=250\mpc$, where the pixel size of $3.91\mpc$ is deemed adequate for the
smoothing length of $6\mpc$.  Next, we smooth the galaxy density field by
convolving it with a Gaussian window function (Eq.~\ref{window_function}) at
$\lambda=6\mpc$.  Finally, for each sample we compute the genus at $100$
values of $\nu$, where $\nu$ is defined in terms of the contour's enclosed
volume fraction (Eqn.~\ref{vfraction}).

In Fig.~\ref{fig:cic}, we show the resulting genus curves for the CIC method
and a Gaussian smoothing with $\lambda=6\mpc$, for different galaxy
catalogues: based on the CLF approach (upper row panels), the SAM of
\citet{Croton2006} (middle row panels), and the SAM of \citet{Bower2006}
(lower row panels), respectively. The panels in the left to right columns give
the results for galaxies with absolute magnitude within $-21.7< \rmag <-20.1$,
$-19.8< \rmag <-19.2$, and $-18.8< \rmag <-18.3$, respectively.  In each
panel, the mean genus curve of the $8$ subsamples is depicted by the solid
line, while the shaded areas indicates the $1\sigma$ scatter calculated from
the $8$ subsamples.  Compared to the Gaussian fit, which is shown as the
dotted line, the genus measurements indicate that after CIC assignment and
Gaussian smoothing with $\lambda=6\mpc$, the galaxy density fields are
consistent with a Gaussian distribution at slightly more than $1\sigma$ level.

One step further, we follow \citet{Park1992} to measure the shift parameter
$\Delta\nu$ of the genus curves, as listed in each panel of
Fig.~\ref{fig:cic}.  In the bright absolute magnitude bin $-21.7< \rmag
<-20.1$, we find that the CLF and Croton galaxy catalogues have positive
$\Delta\nu >0$ at about a $2\sigma$ significance, which means that void
structures dominate the galaxy distribution.  \citet{Gott2008} report a value
of $\Delta\nu=0.010\pm0.023$ for mock samples from Millennium Simulation,
which is in agreement with our measurements for galaxies in different absolute
magnitude bins and in different galaxy catalogues.
 
Following \citet{Park2005a}, we also measured the void multiplicity parameter
$A_V$ and the cluster multiplicity parameter $A_C$, respectively. As can be
seen in Fig.~\ref{fig:cic}, the void multiplicity parameter $A_V$ is lower
than $1$ in all the measurements, especially for galaxies in the bright
absolute magnitude bin $-21.7< \rmag <-20.1$, which implies that voids are
very empty and coalesce into fewer large voids than would be expected for a
Gaussian field. The value $A_V$ slightly rises for fainter galaxies in all the
three galaxy catalogues.  On the other hand, however, the cluster multiplicity
parameter $A_C$ is quite different among the three galaxy catalogues.  It is
consistently below unity in the CLF galaxy catalogue, indicating that there
are fewer independent isolated high density regions than for a Gaussian random
field.  In the \citet{Croton2006} galaxy catalogue in the other hand, there is
a trend that the values of $A_C$ rise and are consistent with unity only for
fainter galaxies.  Overall, however, we find that the amplitude of $A_C$ for
the \citet{Bower2006} galaxy catalogue is consistent with unity. Also, $A_V$
shows smaller deviations than found in the other two catalogues, suggesting
that the galaxy distribution in Bower et al.~indeed shows subtle differences
from the other two.

\subsubsection{Genus Curve for an adaptive smoothing kernel}

We next turn to an analysis of the genus based on adaptively constructed
density fields, considering first the `classic' adaptive kernel estimation
technique and in the next subsection our new tessellation method that works
directly with the point set.

For each of the $8$ subsamples of three galaxy catalogues, we compute the
density field on a regular $64^3$ grid using the SPH method with a neighbor
number of $N_{\rm sph}=64$ galaxies.  Then we calculate the genus curve at
$100$ values of $\nu$. In the middle panel of Fig.~\ref{fig:method}, we show
the genus curves based on the SPH method for our three theoretical galaxy
catalogues in the magnitude range $-21.7< \rmag <-20.1$.

Note first that the number of structural elements resolved with the SPH
adaptive smoothing is much larger than the one accessible with fixed
smoothing.  Compared with the $\lambda=6\mpc$ fixed smoothing results, which
are shown in the left panel of Fig.~\ref{fig:method}, we find that the $N_{\rm
  sph}=64$ adaptive smoothing scheme reaches a genus density which is
approximately three times larger, which is due to the considerably better
adaptive resolution of the corresponding isodensity surfaces.  However,
similar to the fixed smoothing method, the results of the genus statistics for
the three catalogues are clearly well consistent with each other, lacking the
power of discriminating between the different assumptions about the galaxy
formation physics made in the models.

\subsubsection{Genus Curve for the DTFE technique}

Finally, we consider our new DTFE method for density reconstruction and
measure the genus statistics directly on the Delaunay tessellation defined by
the galaxy coordinates.  This method is free of any smoothing parameter, and
should be able to resolve the largest number of structural elements.  The
results are shown in the right panel of Fig.~\ref{fig:method}, again for three
different galaxy catalogues in the magnitude range $-21.7< \rmag <-20.1$.  The
squares with error bars, connected with a solid line, show the results for our
primary CLF catalogue.  The dotted and dashed lines give the corresponding
results for the other two catalogues.  Overall, the genus curve shows a much
larger amplitude than can be resolved with the adaptive smoothing, which is a
direct consequence of the larger resolving power of this method.
Interestingly, we find that there is a significant genus differences between
the three galaxy catalogues, especially between the first two and the SAM of
\citet{Bower2006}. This directly demonstrates the improved discriminative
power made possible by the DTFE genus approach, which here can pick up the
subtle differences introduced in the galaxy distribution due to different
assumptions made in the theoretical galaxy formation modeling.

Apart from the different implementation of galaxy formation processes that
cause the different halo occupation number distribution for galaxies of given
luminosity, e.g., with $-21.7< \rmag <-20.1$, we consider galaxies of
different luminosity ranges.  As explicitly modeled in the CLF approach
introduced by \citet{Yang2003}, galaxies with different luminosity may be
hosted by halos of different masses, which in turn can have quite different
genus behavior. According to the halo mass distribution of galaxies with given
luminosity shown in Fig.~3 of \citet{Yang2009}, we roughly expect that
galaxies with absolute magnitude in the range $-21.7< \rmag <-20.1$ should be
hosted by halos of mass $\sim 10^{12}\msun$ if they are central galaxies, and
in larger halos if they are satellite galaxies. Galaxies with $-19.8< \rmag
<-19.2$ and $-18.8< \rmag <-18.3$ should be hosted in halos of mass $\sim
10^{11.6}\msun$ and $\sim 10^{11.3}\msun$, respectively, as central galaxies
and in larger halos as satellite galaxies.  We show and compare in each panel
of Fig.~\ref{fig:magnitude} the genus statistics of galaxies in these three
different magnitude ranges.  Results shown in the different panels correspond
to different galaxy catalogues (left to right columns) and different genus
measurement methods (upper to lower rows).  As can be seen, only in the three
bottom panels based on the DTFE method can we clearly distinguish the
difference among the genus curves for galaxies in different magnitude ranges,
at much better than a $1\sigma$ level.  Note also, in the DTFE results, the
\citet{Bower2006} galaxy catalogue reveals significantly smaller differences
between galaxies within the different absolute magnitude ranges than the other
catalogues.  This behavior can be well understood if we check the halo mass
distribution of these galaxies: for a given luminosity, the \citet{Bower2006}
model shows comparatively large halo mass scatter.  And thus in
\citet{Bower2006} there is a larger overlap of the halo masses in the three
different magnitude ranges (see also Fig. 4 in Liu et al. 2010).

\subsection{Results for galaxies in redshift space}

\begin{figure*}
\begin{center}
\resizebox{18cm}{!}{\includegraphics{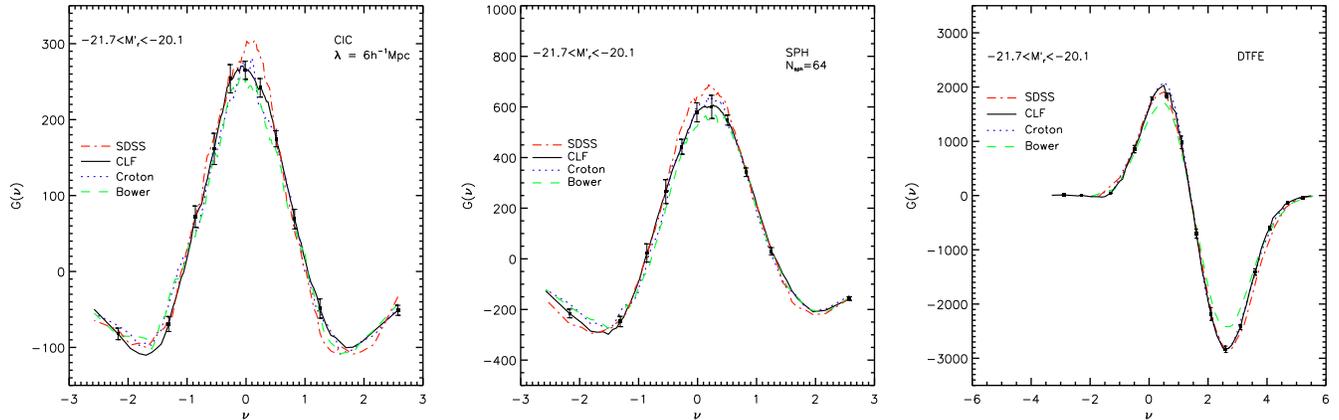}}\vspace*{-0.5cm}
\end{center}
\caption{Comparison of the genus statistics between the SDSS and mock redshift
  surveys that take into account various observational effects, especially the
  redshift distortion effect, using different density reconstruction methods
  from left to right, as indicated in the panels.  The error bars are
  calculated from $6$ mock redshift surveys (with different pointings) based
  on the galaxy catalogue constructed using the conditional luminosity
  function.}
\label{fig:sdss}
\end{figure*}

\begin{figure}
\begin{center}
\resizebox{8cm}{!}{\includegraphics{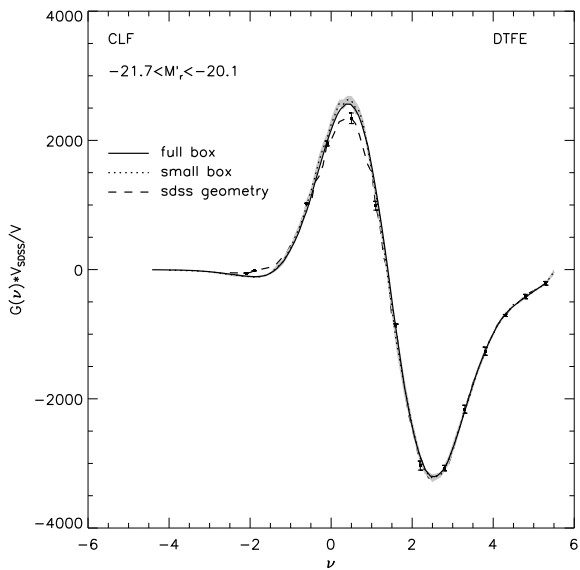}}
\end{center}
\caption{Comparison of the genus statistics for the DTFE method
  measured from the full simulation box (solid line), from a smaller
  box of equal volume to the SDSS (dotted line), and from the SDSS
  geometry (dashed line with error bars) for the CLF galaxy catalogue
  based on the Millennium Simulation. The results shown in this plot
  are all measured in real space.  The error bars for the SDSS
  geometry results are again calculated from $6$ mock redshift surveys
  (with different pointings) based on the CLF catalogue, but here in
  real space. The shaded area is the standard error of 8 subsamples.}
\label{fig:geometry}
\end{figure}

\begin{figure}
\begin{center}
\resizebox{8cm}{!}{\includegraphics{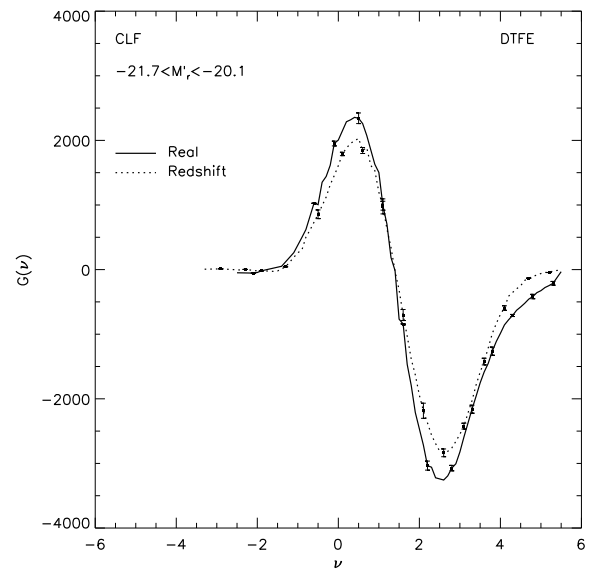}}
\end{center}
\caption{Comparison of the genus statistics for the DTFE method in
  real (solid line) and redshift (dotted line) space for MGRSs based
  on the CLF galaxy catalogue. The results in real and redshift spaces
  are the same as those shown in Fig.~\ref{fig:geometry} and the right
  panel of Fig.~\ref{fig:sdss}, respectively. }
\label{fig:distortion}
\end{figure}

Having measured the genus statistics in real space using different methods for
three different galaxy catalogues and for galaxies in three different
luminosity ranges, we now turn to an analysis where observational selection
effects are included.  This also allows us to make a direct comparison with
the SDSS observations in an ``apples-to-apples'' fashion. We note however that
we here compare only to theoretical predictions obtained for the $\Lambda$CDM
cosmology simulated in the Millennium Simulation; in the future, it will be
very interesting to also carry out such a comparison for different
cosmological models. Also we remark that although we have demonstrated in the
previous subsections that the genus measured with the DTFE method can in
principle distinguish topological differences between different galaxy
catalogues constructed for the same simulation model, and galaxies within
different absolute magnitude ranges, we lack the theoretical
prediction/judgment of which of these models is ultimately the most accurate.

As an illustration, we here use the SDSS observation to judge to what extent
the three galaxy catalogues we discussed in this paper agree with the
observations.  Note that in SDSS, there are a number of important survey
selection effects: flux limit, sky mask, redshift distortion, etc. To have a
fair comparison between the observations and the galaxies in the three
theoretical catalogues, we have generated appropriate mock galaxy redshift
surveys (MGRSs). We then compare the genus statistics measured for the
volume-limited samples with $-21.7< \rmag <-20.1$ extracted from the SDSS
observations and for the MGRSs corresponding to our three theoretical galaxy
catalogues.  The main differences between these measurements and the ones
carried out earlier in cubic boxes are: (1) a light cone geometry without
periodic boundary conditions is used, and (2) the analysis is in redshift
space.

In Fig.~\ref{fig:sdss}, we show the genus curves of SDSS and the MGRSs
generated from the three different galaxy catalogues.  Results shown in the
left, middle and right panels are for genus statistics measured using the
$\lambda=6\mpc$ fixed smoothing, the SPH $N_{\rm sph}=64$ adaptive smoothing,
and the DTFE method, respectively. The error bars on top of the CLF results
are obtained from $1\sigma$ variances of 6 MGRSs generated from the CLF galaxy
catalogue by adopting different pointings. As one can see from the plot, the
first two genus measurement methods indicate that the three galaxy catalogues
are consistent with the SDSS observations at about a 1-2$\sigma$ level, with
the \citet{Bower2006} model doing slightly worse. However, the DTFE method
clearly shows that the \citet{Bower2006} galaxy catalogue is actually
inconsistent with the SDSS observations at a high significance level
\citep[see also][]{Choi2010}.  Again, this is a strong demonstration of the
improved discriminative power of the DTFE genus measurement technique even
when used in redshift space.

  Finally, we examine the systematic effects that our SDSS survey
  selections may induce. For illustration purposes, we focus only on
  the DTFE technique and check the impacts of the light cone geometry
  and of redshift space distortions. We first investigate the light
  cone geometry effect (according to the SDSS mask) on the shape and
  amplitude of the genus curve. In Fig.~\ref{fig:geometry}, we show
  three DTFE genus curves based on the CLF galaxy catalogue, scaled to
  the same volume: (i) computed for the cubic box of $L_{\rm
    box}=500\mpc$, (ii) for small boxes of equal volume as the SDSS,
  and (iii) for MGRSs selected with exactly the same SDSS geometry.
  Note that all the results shown in this plot are measured in real
  space.  The small and large box results are in overall good
  agreement at better than 1-$\sigma$ level. However, the results for
  the SDSS geometry do suffer somewhat from the edge effects, biasing
  the genus curve to a lower (by about 2-$\sigma$) maximum value.

  Next we check the impact of redshift space distortions. In
  Fig.~\ref{fig:distortion}, we compare the DTFE genus curves for the
  MGRSs based on the CLF galaxy catalogue, measured separately in real
  and redshift spaces. Both the maximum and minimum values of the
  genus curves in real space are more prominent than those in redshift
  space, and their shapes are also slightly different, emphasizing the
  need of making ``apples-to-apples'' comparisons with the SDSS
  observations when genus measurements are considered. With further
  tests we have confirmed that similar differences also exist in the
  fixed and adaptive smoothing genus measurements.

In this paper, we have focused on introducing the new DTFE methodology and
showing its potential. We therefore restricted our analysis to a comparison of
the genus statistics of SDSS observations with galaxy models generated for a
single simulation model, the Millennium Simulation, using only $\sim L^{\ast}$
galaxies.  We defer a more detailed comparison with the SDSS observations to a
forthcoming paper, where also galaxies generated in simulations with different
cosmological parameters, and galaxies in different luminosity bins, etc., are
studied. This will also further clarify the ability of the DTFE genus
technique to provide constraints on cosmological and galaxy formation
parameters.

\section{CONCLUSIONS}\label{sec_conclusions}

In this study, we introduced a new method for measuring the genus statistics
of isodensity surfaces of the galaxy distribution, based on the Delaunay
tessellation field estimation for constructing a continuous and piece-wise
linear galaxy density field. Compared to traditional Gaussian smoothing and
adaptive SPH smoothing, this technique does not require any free parameter,
and it allows the extraction of the largest amount of topological information
from the point set.

To demonstrate the abilities of the new method for measuring the topology of
the large-scale structure compared to the traditional methods, we have carried
out genus measurements using galaxies both in real space and in redshift
space. In real space, we make use of three galaxy catalogues constructed for
the same underlying Millennium Simulation: one based on the CLF approach and
the other two generated with semi-analytical galaxy formation models. In
redshift space, we make use of the volume-limited samples extracted from the
SDSS observational data and the MGRSs generated using the three galaxy
catalogues.  Three types of genus measurement methods are introduced and
applied to those samples: (i) CIC assignment with $\lambda=6\mpc$ fixed
Gaussian smoothing; (ii) adaptive SPH-like smoothing with a neighbor number of
$N_{\rm sph}=64$ galaxies; and (iii) Delaunay tessellation field estimation.

Based on our various comparisons, we can summarize our findings as follows.
\begin{itemize}
\item The traditional genus measurement method based on a fixed smoothing can
  not really distinguish the topology of the different galaxy formation models
  considered here. Reassuringly, they are all found to be consistent with the
  SDSS observations. If adaptive smoothing is used instead, more topological
  information can be recovered, but only marginal differences between the
  models at a 1-2$\sigma$ level can be detected.
\item The DTFE method has significantly enhanced genus measurement power: its
  larger amplitude in $G(\nu)$ reflects its ability to extract a maximum
  amount of topological information from the galaxy density field.
\item Most importantly, the DTFE method can distinguish the topology of
  different galaxy formation models and of galaxies in different luminosity
  ranges at very significant confidence levels, both in real and in redshift
  space.
\item Comparing with the SDSS observational data using MGRSs that take into
  account various survey selection effects, we find that the semi-analytical
  galaxy catalogue constructed by \citet{Bower2006} deviates from the
  observations significantly.
\end{itemize}

Our results have clearly demonstrated the power of our new DTFE method in
performing a topological analysis both in real and in redshift space.  The
present work is the first study in a series of papers where we want to exploit
the full potential of the new approach for constraining cosmological
parameters and the galaxy formation process. It will also be interesting to
see how sensitive the DTFE method is to traces of non-Gaussianity in the
initial conditions, or to non-standard dark energy models.

\acknowledgements

We  thank  Changbom  Park  \&  Yun-Young  Choi for  kindly  providing  us  the
KIAS-VAGC, and the  anonymous referee for useful and  insightful comments that
greatly helped to improve the presentation of this paper.  This work is partly
supported  by 973 Program  (No.  2007CB815402),  the CAS  Knowledge Innovation
Program  (Grant  No.  KJCX2-YW-T05)  and  grants  from  NSFC (Nos.   10821302,
10925314).

The Millennium  Simulation used  in this  paper was carried  out by  the Virgo
Supercomputing Consortium at the Computing Centre of the Max-Planck Society in
Garching.  The  semi-analytic  galaxy   catalogue  is  publicly  available  at
http://www.mpa-garching.mpg.de/galform/agnpaper.

Funding for  the SDSS  and SDSS-II has  been provided  by the Alfred  P. Sloan
Foundation, the  Participating Institutions, the  National Science Foundation,
the   U.S.  Department  of   Energy,  the   National  Aeronautics   and  Space
Administration, the  Japanese Monbukagakusho, the Max Planck  Society, and the
Higher  Education  Funding   Council  for  England.  The  SDSS   Web  Site  is
http://www.sdss.org/.

The  SDSS  is  managed  by  the  Astrophysical  Research  Consortium  for  the
Participating  Institutions. The Participating  Institutions are  the American
Museum  of Natural  History,  Astrophysical Institute  Potsdam, University  of
Basel, University of Cambridge, Case Western Reserve University, University of
Chicago, Drexel  University, Fermilab, the  Institute for Advanced  Study, the
Japan Participation  Group, Johns Hopkins University, the  Joint Institute for
Nuclear  Astrophysics,  the  Kavli  Institute for  Particle  Astrophysics  and
Cosmology,  the  Korean  Scientist  Group,  the Chinese  Academy  of  Sciences
(LAMOST),  Los  Alamos   National  Laboratory,  the  Max-Planck-Institute  for
Astronomy (MPIA), the Max-Planck-Institute  for Astrophysics (MPA), New Mexico
State University, Ohio State  University, University of Pittsburgh, University
of Portsmouth, Princeton University,  the United States Naval Observatory, and
the University of Washington.


\begin{thebibliography}{54}
\expandafter\ifx\csname natexlab\endcsname\relax\def\natexlab#1{#1}\fi

\bibitem[{{Abazajian} {et~al.}(2009){Abazajian}, {Adelman-McCarthy},
  {Ag{\"u}eros}, {Allam}, {Allende Prieto}, {An}, {Anderson}, {Anderson},
  {Annis}, {Bahcall}, {Bailer-Jones}, {Barentine}, {Bassett}, {Becker},
  {Beers}, {Bell}, {Belokurov}, {Berlind}, {Berman}, {Bernardi}, {Bickerton},
  {Bizyaev}, {Blakeslee}, {Blanton}, {Bochanski}, {Boroski}, {Brewington},
  {Brinchmann}, {Brinkmann}, {Brunner}, {Budav{\'a}ri}, {Carey}, {Carliles},
  {Carr}, {Castander}, {Cinabro}, {Connolly}, {Csabai}, {Cunha}, {Czarapata},
  {Davenport}, {de Haas}, {Dilday}, {Doi}, {Eisenstein}, {Evans}, {Evans},
  {Fan}, {Friedman}, {Frieman}, {Fukugita}, {G{\"a}nsicke}, {Gates},
  {Gillespie}, {Gilmore}, {Gonzalez}, {Gonzalez}, {Grebel}, {Gunn},
  {Gy{\"o}ry}, {Hall}, {Harding}, {Harris}, {Harvanek}, {Hawley}, {Hayes},
  {Heckman}, {Hendry}, {Hennessy}, {Hindsley}, {Hoblitt}, {Hogan}, {Hogg},
  {Holtzman}, {Hyde}, {Ichikawa}, {Ichikawa}, {Im}, {Ivezi{\'c}}, {Jester},
  {Jiang}, {Johnson}, {Jorgensen}, {Juri{\'c}}, {Kent}, {Kessler}, {Kleinman},
  {Knapp}, {Konishi}, {Kron}, {Krzesinski}, {Kuropatkin}, {Lampeitl},
  {Lebedeva}, {Lee}, {Lee}, {Leger}, {L{\'e}pine}, {Li}, {Lima}, {Lin}, {Long},
  {Loomis}, {Loveday}, {Lupton}, {Magnier}, {Malanushenko}, {Malanushenko},
  {Mandelbaum}, {Margon}, {Marriner}, {Mart{\'{\i}}nez-Delgado}, {Matsubara},
  {McGehee}, {McKay}, {Meiksin}, {Morrison}, {Mullally}, {Munn}, {Murphy},
  {Nash}, {Nebot}, {Neilsen}, {Newberg}, {Newman}, {Nichol}, {Nicinski},
  {Nieto-Santisteban}, {Nitta}, {Okamura}, {Oravetz}, {Ostriker}, {Owen},
  {Padmanabhan}, {Pan}, {Park}, {Pauls}, {Peoples}, {Percival}, {Pier}, {Pope},
  {Pourbaix}, {Price}, {Purger}, {Quinn}, {Raddick}, {Fiorentin}, {Richards},
  {Richmond}, {Riess}, {Rix}, {Rockosi}, {Sako}, {Schlegel}, {Schneider},
  {Scholz}, {Schreiber}, {Schwope}, {Seljak}, {Sesar}, {Sheldon}, {Shimasaku},
  {Sibley}, {Simmons}, {Sivarani}, {Smith}, {Smith}, {Smol{\v c}i{\'c}},
  {Snedden}, {Stebbins}, {Steinmetz}, {Stoughton}, {Strauss}, {Subba Rao},
  {Suto}, {Szalay}, {Szapudi}, {Szkody}, {Tanaka}, {Tegmark}, {Teodoro},
  {Thakar}, {Tremonti}, {Tucker}, {Uomoto}, {Vanden Berk}, {Vandenberg},
  {Vidrih}, {Vogeley}, {Voges}, {Vogt}, {Wadadekar}, {Watters}, {Weinberg},
  {West}, {White}, {Wilhite}, {Wonders}, {Yanny}, {Yocum}, {York}, {Zehavi},
  {Zibetti}, \& {Zucker}}]{Abazajian2009}
{Abazajian}, K.~N., {et~al.} 2009, \apjs, 182, 543

\bibitem[{{Barnes} \& {Hut}(1986)}]{Barnes1986}
{Barnes}, J., \& {Hut}, P. 1986, \nat, 324, 446

\bibitem[{{Blanton} {et~al.}(2003){Blanton}, {Hogg}, {Bahcall}, {Brinkmann},
  {Britton}, {Connolly}, {Csabai}, {Fukugita}, {Loveday}, {Meiksin}, {Munn},
  {Nichol}, {Okamura}, {Quinn}, {Schneider}, {Shimasaku}, {Strauss}, {Tegmark},
  {Vogeley}, \& {Weinberg}}]{Blanton2003}
{Blanton}, M.~R., {et~al.} 2003, \apj, 592, 819

\bibitem[{{Blanton} {et~al.}(2005){Blanton}, {Schlegel}, {Strauss},
  {Brinkmann}, {Finkbeiner}, {Fukugita}, {Gunn}, {Hogg}, {Ivezi{\'c}}, {Knapp},
  {Lupton}, {Munn}, {Schneider}, {Tegmark}, \& {Zehavi}}]{Blanton2005}
---. 2005, \aj, 129, 2562

\bibitem[{{Bower} {et~al.}(2006){Bower}, {Benson}, {Malbon}, {Helly}, {Frenk},
  {Baugh}, {Cole}, \& {Lacey}}]{Bower2006}
{Bower}, R.~G., {Benson}, A.~J., {Malbon}, R., {Helly}, J.~C., {Frenk}, C.~S.,
  {Baugh}, C.~M., {Cole}, S., \& {Lacey}, C.~G. 2006, \mnras, 370, 645

\bibitem[{{Cacciato} {et~al.}(2009){Cacciato}, {van den Bosch}, {More}, {Li},
  {Mo}, \& {Yang}}]{Cacciato2009}
{Cacciato}, M., {van den Bosch}, F.~C., {More}, S., {Li}, R., {Mo}, H.~J., \&
  {Yang}, X. 2009, \mnras, 394, 929

\bibitem[{{Canavezes} {et~al.}(1998){Canavezes}, {Springel}, {Oliver},
  {Rowan-Robinson}, {Keeble}, {White}, {Saunders}, {Efstathiou}, {Frenk},
  {McMahon}, {Maddox}, {Sutherland}, \& {Tadros}}]{Canavezes1998}
{Canavezes}, A., {et~al.} 1998, \mnras, 297, 777

\bibitem[{{Choi} {et~al.}(2010){Choi}, {Park}, {Kim}, {Gott}, {Weinberg},
  {Vogeley}, \& {Kim}}]{Choi2010}
{Choi}, Y., {Park}, C., {Kim}, J., {Gott}, III, J.~R., {Weinberg}, D.~H.,
  {Vogeley}, M.~S., \& {Kim}, S.~S. 2010, ArXiv:1005.0256

\bibitem[{{Choi} {et~al.}(2007){Choi}, {Park}, \& {Vogeley}}]{Choi2007}
{Choi}, Y., {Park}, C., \& {Vogeley}, M.~S. 2007, \apj, 658, 884

\bibitem[{{Colless} {et~al.}(2001){Colless}, {Dalton}, {Maddox}, {Sutherland},
  {Norberg}, {Cole}, {Bland-Hawthorn}, {Bridges}, {Cannon}, {Collins}, {Couch},
  {Cross}, {Deeley}, {De Propris}, {Driver}, {Efstathiou}, {Ellis}, {Frenk},
  {Glazebrook}, {Jackson}, {Lahav}, {Lewis}, {Lumsden}, {Madgwick}, {Peacock},
  {Peterson}, {Price}, {Seaborne}, \& {Taylor}}]{Colless2001}
{Colless}, M., {et~al.} 2001, \mnras, 328, 1039

\bibitem[{{Croton} {et~al.}(2006){Croton}, {Springel}, {White}, {De Lucia},
  {Frenk}, {Gao}, {Jenkins}, {Kauffmann}, {Navarro}, \& {Yoshida}}]{Croton2006}
{Croton}, D.~J., {et~al.} 2006, \mnras, 365, 11

\bibitem[{{Dave} {et~al.}(1997){Dave}, {Hellinger}, {Primack}, {Nolthenius}, \&
  {Klypin}}]{Dave1997}
{Dave}, R., {Hellinger}, D., {Primack}, J., {Nolthenius}, R., \& {Klypin}, A.
  1997, \mnras, 284, 607

\bibitem[{{Davis} {et~al.}(1985){Davis}, {Efstathiou}, {Frenk}, \&
  {White}}]{Davis1985}
{Davis}, M., {Efstathiou}, G., {Frenk}, C.~S., \& {White}, S.~D.~M. 1985, \apj,
  292, 371

\bibitem[{{De Lucia} \& {Blaizot}(2007)}]{Delucia2007}
{De Lucia}, G., \& {Blaizot}, J. 2007, \mnras, 375, 2

\bibitem[{Delaunay(1934)}]{Delaunay1934}
Delaunay, B.~N. 1934, Sur la sph\`ere vide: Izv. Akad. Nauk SSSR, Otdel. Mat.
  Est. Nauk, 7, 793

\bibitem[{{Dunkley} {et~al.}(2009){Dunkley}, {Komatsu}, {Nolta}, {Spergel},
  {Larson}, {Hinshaw}, {Page}, {Bennett}, {Gold}, {Jarosik}, {Weiland},
  {Halpern}, {Hill}, {Kogut}, {Limon}, {Meyer}, {Tucker}, {Wollack}, \&
  {Wright}}]{Dunkley2009}
{Dunkley}, J., {et~al.} 2009, \apjs, 180, 306

\bibitem[{{Gott} {et~al.}(2009){Gott}, {Choi}, {Park}, \& {Kim}}]{Gott2009}
{Gott}, J.~R., {Choi}, Y., {Park}, C., \& {Kim}, J. 2009, \apjl, 695, L45

\bibitem[{{Gott} {et~al.}(1986){Gott}, {Dickinson}, \& {Melott}}]{Gott1986}
{Gott}, III, J.~R., {Dickinson}, M., \& {Melott}, A.~L. 1986, \apj, 306, 341

\bibitem[{{Gott} {et~al.}(1987){Gott}, {Weinberg}, \& {Melott}}]{Gott1987}
{Gott}, J.~R.~I., {Weinberg}, D.~H., \& {Melott}, A.~L. 1987, \apj, 319, 1

\bibitem[{{Gott} {et~al.}(1989){Gott}, {Miller}, {Thuan}, {Schneider},
  {Weinberg}, {Gammie}, {Polk}, {Vogeley}, {Jeffrey}, {Bhavsar}, {Melott},
  {Giovanelli}, {Hayes}, {Tully}, \& {Hamilton}}]{Gott1989}
{Gott}, J.~R.~I., {et~al.} 1989, \apj, 340, 625

\bibitem[{{Gott} {et~al.}(2008){Gott}, {Hambrick}, {Vogeley}, {Kim}, {Park},
  {Choi}, {Cen}, {Ostriker}, \& {Nagamine}}]{Gott2008}
---. 2008, \apj, 675, 16

\bibitem[{{Guth}(1981)}]{Guth1981}
{Guth}, A.~H. 1981, \prd, 23, 347

\bibitem[{{Guth} \& {Pi}(1982)}]{Guth1982}
{Guth}, A.~H., \& {Pi}, S. 1982, Physical Review Letters, 49, 1110

\bibitem[{{Hamilton} {et~al.}(1986){Hamilton}, {Gott}, \&
  {Weinberg}}]{Hamilton1986}
{Hamilton}, A.~J.~S., {Gott}, J.~R.~I., \& {Weinberg}, D. 1986, \apj, 309, 1

\bibitem[{{Hernquist} \& {Katz}(1989)}]{Hernquist1989}
{Hernquist}, L., \& {Katz}, N. 1989, \apjs, 70, 419

\bibitem[{{Hikage} {et~al.}(2002){Hikage}, {Suto}, {Kayo}, {Taruya},
  {Matsubara}, {Vogeley}, {Hoyle}, {Gott}, \& {Brinkmann}}]{Hikage2002}
{Hikage}, C., {et~al.} 2002, \pasj, 54, 707

\bibitem[{{Hikage} {et~al.}(2003){Hikage}, {Schmalzing}, {Buchert}, {Suto},
  {Kayo}, {Taruya}, {Vogeley}, {Hoyle}, {Gott}, \& {Brinkmann}}]{Hikage2003}
---. 2003, \pasj, 55, 911

\bibitem[Illian et al. (2008)]{Illian2008} Illian, J., Penttinen, A., Stoyan,
  H., and Stoyan, D. 2008, Statistical Analysis and Modelling of Spatial Point
  Patterns. (Wiley-Interscience).


\bibitem[{{James} {et~al.}(2009){James}, {Colless}, {Lewis}, \&
  {Peacock}}]{James2009}
{James}, J.~B., {Colless}, M., {Lewis}, G.~F., \& {Peacock}, J.~A. 2009,
  \mnras, 394, 454

\bibitem[{{Kofman} {et~al.}(1994){Kofman}, {Bertschinger}, {Gelb}, {Nusser}, \&
  {Dekel}}]{Kofman1994}
{Kofman}, L., {Bertschinger}, E., {Gelb}, J.~M., {Nusser}, A., \& {Dekel}, A.
  1994, \apj, 420, 44

\bibitem[{{Li} {et~al.}(2007){Li}, {Jing}, {Kauffmann}, {B{\"o}rner}, {Kang},
  \& {Wang}}]{Li2007}
{Li}, C., {Jing}, Y.~P., {Kauffmann}, G., {B{\"o}rner}, G., {Kang}, X., \&
  {Wang}, L. 2007, \mnras, 376, 984

\bibitem[{{Liu} {et~al.}(2010){Liu}, {Yang}, {Mo}, {van den Bosch}, \&
  {Springel}}]{Liu2010}
{Liu}, L., {Yang}, X., {Mo}, H.~J., {van den Bosch}, F.~C., \& {Springel}, V.
  2010, \apj, 712, 734

\bibitem[{{Mecke} {et~al.}(1994){Mecke}, {Buchert}, \& {Wagner}}]{Mecke1994}
{Mecke}, K.~R., {Buchert}, T., \& {Wagner}, H. 1994, \aap, 288, 697

\bibitem[{{Monaghan} \& {Lattanzio}(1985)}]{Monaghan1985}
{Monaghan}, J.~J., \& {Lattanzio}, J.~C. 1985, \aap, 149, 135

\bibitem[{{Moore} {et~al.}(1992){Moore}, {Frenk}, {Weinberg}, {Saunders},
  {Lawrence}, {Ellis}, {Kaiser}, {Efstathiou}, \& {Rowan-Robinson}}]{Moore1992}
{Moore}, B., {et~al.} 1992, \mnras, 256, 477

\bibitem[{Okabe} {et~al}(2000)]{Okabe2000} Okabe, A. {et~al}\ 2000, Spatial
  tessellations : concepts and applications of voronoi diagrams. ~2nd
  ed. Chichester: John Wiley， 2000， 1-671

\bibitem[{{Park} \& {Choi}(2005)}]{Park2005c}
{Park}, C., \& {Choi}, Y. 2005, \apjl, 635, L29

\bibitem[{{Park} {et~al.}(1992){Park}, {Gott}, \& {da Costa}}]{Park1992}
{Park}, C., {Gott}, III, J.~R., \& {da Costa}, L.~N. 1992, \apjl, 392, L51

\bibitem[{{Park} {et~al.}(2005){Park}, {Kim}, \& {Gott}}]{Park2005a}
{Park}, C., {Kim}, J., \& {Gott}, J.~R.~I. 2005, \apj, 633, 1

\bibitem[{{Peebles}(1980)}]{Peebles1980}
{Peebles}, P.~J.~E. 1980, {The large-scale structure of the universe},
  Princeton, N.J., Princeton Univ. Press

\bibitem[{{Pelupessy} {et~al.}(2003){Pelupessy}, {Schaap}, \& {van de
  Weygaert}}]{Pelupessy2003}
{Pelupessy}, F.~I., {Schaap}, W.~E., \& {van de Weygaert}, R. 2003, \aap, 403,
  389

\bibitem[{{Protogeros} \& {Weinberg}(1997)}]{Protogeros1997}
{Protogeros}, Z.~A.~M., \& {Weinberg}, D.~H. 1997, \apj, 489, 457

\bibitem[{{Rhoads} {et~al.}(1994){Rhoads}, {Gott}, \& {Postman}}]{Rhoads1994}
{Rhoads}, J.~E., {Gott}, J.~R.~I., \& {Postman}, M. 1994, \apj, 421, 1

\bibitem[{{Schaap} \& {van de Weygaert}(2000)}]{Schaap2000}
{Schaap}, W.~E., \& {van de Weygaert}, R. 2000, \aap, 363, L29

\bibitem[{{Springel}(2005)}]{Springel2005a}
{Springel}, V. 2005, \mnras, 364, 1105

\bibitem[{{Springel}(2009)}]{Springel2009}
---. 2009, \mnras, 1655

\bibitem[{{Springel} {et~al.}(2006){Springel}, {Frenk}, \&
  {White}}]{Springel2006}
{Springel}, V., {Frenk}, C.~S., \& {White}, S.~D.~M. 2006, \nat, 440, 1137

\bibitem[{{Springel} {et~al.}(2001){Springel}, {White}, {Tormen}, \&
  {Kauffmann}}]{Springel2001a}
{Springel}, V., {White}, S.~D.~M., {Tormen}, G., \& {Kauffmann}, G. 2001,
  \mnras, 328, 726

\bibitem[{{Springel} {et~al.}(1998){Springel}, {White}, {Colberg}, {Couchman},
  {Efstathiou}, {Frenk}, {Jenkins}, {Pearce}, {Nelson}, {Peacock}, \&
  {Thomas}}]{Springel1998}
{Springel}, V., {et~al.} 1998, \mnras, 298, 1169

\bibitem[{{Springel} {et~al.}(2005){Springel}, {White}, {Jenkins}, {Frenk},
  {Yoshida}, {Gao}, {Navarro}, {Thacker}, {Croton}, {Helly}, {Peacock}, {Cole},
  {Thomas}, {Couchman}, {Evrard}, {Colberg}, \& {Pearce}}]{Springel2005b}
---. 2005, \nat, 435, 629

\bibitem[{{Vogeley} {et~al.}(1994){Vogeley}, {Park}, {Geller}, {Huchra}, \&
  {Gott}}]{Vogeley1994}
{Vogeley}, M.~S., {Park}, C., {Geller}, M.~J., {Huchra}, J.~P., \& {Gott},
  J.~R.~I. 1994, \apj, 420, 525

\bibitem[{{Weinberg}(1988)}]{Weinberg1988}
{Weinberg}, D.~H. 1988, \pasp, 100, 1373

\bibitem[{{Yang} {et~al.}(2004){Yang}, {Mo}, {Jing}, {van den Bosch}, \&
  {Chu}}]{Yang2004}
{Yang}, X., {Mo}, H.~J., {Jing}, Y.~P., {van den Bosch}, F.~C., \& {Chu}, Y.
  2004, \mnras, 350, 1153

\bibitem[{{Yang} {et~al.}(2009){Yang}, {Mo}, \& {van den Bosch}}]{Yang2009}
{Yang}, X., {Mo}, H.~J., \& {van den Bosch}, F.~C. 2009, \apj, 695, 900

\bibitem[{Yang {et~al.}(2003)Yang, Mo, \& van~den Bosch}]{Yang2003}
Yang, X.-h., Mo, H.~J., \& van~den Bosch, F.~C. 2003, Mon. Not. Roy. Astron.
  Soc., 339, 1057

\bibitem[{{York} {et~al.}(2000){York}, {Adelman}, {Anderson}, {Anderson},
  {Annis}, {Bahcall}, {Bakken}, {Barkhouser}, {Bastian}, {Berman}, {Boroski},
  {Bracker}, {Briegel}, {Briggs}, {Brinkmann}, {Brunner}, {Burles}, {Carey},
  {Carr}, {Castander}, {Chen}, {Colestock}, {Connolly}, {Crocker}, {Csabai},
  {Czarapata}, {Davis}, {Doi}, {Dombeck}, {Eisenstein}, {Ellman}, {Elms},
  {Evans}, {Fan}, {Federwitz}, {Fiscelli}, {Friedman}, {Frieman}, {Fukugita},
  {Gillespie}, {Gunn}, {Gurbani}, {de Haas}, {Haldeman}, {Harris}, {Hayes},
  {Heckman}, {Hennessy}, {Hindsley}, {Holm}, {Holmgren}, {Huang}, {Hull},
  {Husby}, {Ichikawa}, {Ichikawa}, {Ivezi{\'c}}, {Kent}, {Kim}, {Kinney},
  {Klaene}, {Kleinman}, {Kleinman}, {Knapp}, {Korienek}, {Kron}, {Kunszt},
  {Lamb}, {Lee}, {Leger}, {Limmongkol}, {Lindenmeyer}, {Long}, {Loomis},
  {Loveday}, {Lucinio}, {Lupton}, {MacKinnon}, {Mannery}, {Mantsch}, {Margon},
  {McGehee}, {McKay}, {Meiksin}, {Merelli}, {Monet}, {Munn}, {Narayanan},
  {Nash}, {Neilsen}, {Neswold}, {Newberg}, {Nichol}, {Nicinski}, {Nonino},
  {Okada}, {Okamura}, {Ostriker}, {Owen}, {Pauls}, {Peoples}, {Peterson},
  {Petravick}, {Pier}, {Pope}, {Pordes}, {Prosapio}, {Rechenmacher}, {Quinn},
  {Richards}, {Richmond}, {Rivetta}, {Rockosi}, {Ruthmansdorfer}, {Sandford},
  {Schlegel}, {Schneider}, {Sekiguchi}, {Sergey}, {Shimasaku}, {Siegmund},
  {Smee}, {Smith}, {Snedden}, {Stone}, {Stoughton}, {Strauss}, {Stubbs},
  {SubbaRao}, {Szalay}, {Szapudi}, {Szokoly}, {Thakar}, {Tremonti}, {Tucker},
  {Uomoto}, {Vanden Berk}, {Vogeley}, {Waddell}, {Wang}, {Watanabe},
  {Weinberg}, {Yanny}, \& {Yasuda}}]{York2000}
{York}, D.~G., {et~al.} 2000, \aj, 120, 1579

\end{thebibliography}
\end{document}